\documentclass[10pt,a4paper]{article}
\usepackage[margin=1in]{geometry}
\usepackage{amsmath,amssymb,amsthm}
\usepackage{tikz}
\usepackage{graphicx}
\usepackage{subcaption}
\usepackage{booktabs}
\usepackage{multirow}
\usepackage{authblk}
\usepackage[hidelinks]{hyperref}
\usepackage[nameinlink]{cleveref}

\newtheorem{theorem}{Theorem}[section]

\theoremstyle{definition}
\newtheorem{definition}[theorem]{Definition}
\newtheorem{claim}[theorem]{Claim}
\theoremstyle{remark}
\newtheorem{remark}[theorem]{Remark}

\makeatletter
\renewcommand\AB@affilsepx{\protect\Affilfont}
\makeatother

\title{Covering Planar Lattices with Interior-Disjoint Unit Disks}

\author{Nattawut Phetmak}
\author{Grittin Nuntasombat}
\author{Jittat Fakcharoenphol}

\affil{\small
  Department of Computer Engineering, \linebreak
  Kasetsart University, Bangkok, Thailand \linebreak
  \texttt{ \{nattawut.p,grittin.n\}@ku.th, jtf@ku.ac.th}
}

\begin{document}

\maketitle
\begin{abstract}
  We study an infinite variant of the coin-covering problem for periodic point sets in the plane. Given a point set of spacing $d$, we ask whether all of its points can be covered by pairwise non-overlapping unit disks. We consider the triangular lattice, the square lattice, and the honeycomb point set, and construct periodic motif patterns that certify several intervals of coverable spacings. For the triangular lattice, our constructions include single-family patterns with vertex, face, and off-lattice realizing centers, as well as multi-family patterns. For the honeycomb point set, additional native motifs fill gaps left by the triangular-lattice constructions. For the square lattice, we revisit the constructions of Alm et~al., identify an unintended overlap in one motif realization, and give new patterns that recover part of the affected interval and establish an additional coverability interval.
\end{abstract}


\begin{figure}[!b]
  \centering
  \input{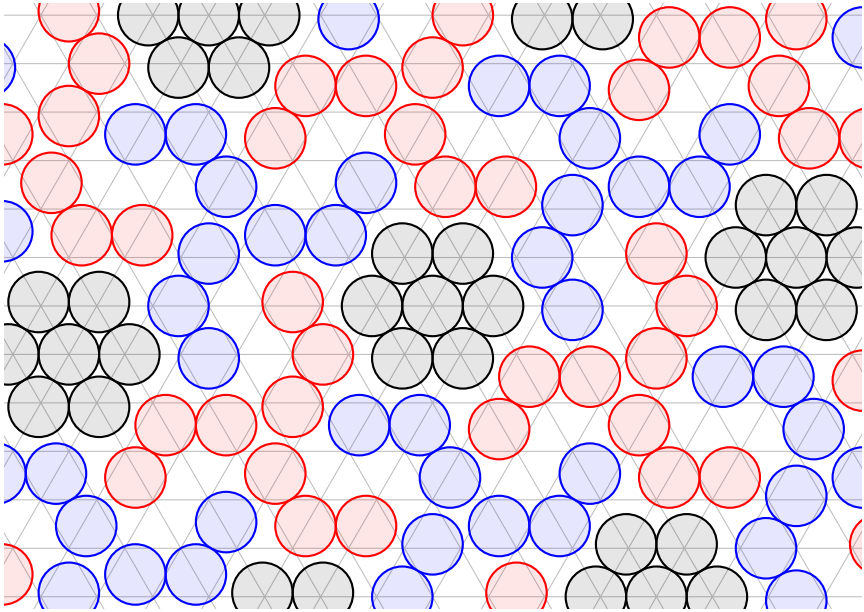}
  \caption{A non-uniform arrangement of coins for covering the triangular lattice.}
  \label{fig:tri-cyclone}
\end{figure}

\section{Introduction}
\label{sec:intro}

The \emph{10-point problem}, proposed by Inaba as a puzzle in 2008, asks the following question: given a collection of identical coins, show that every configuration of 10 points in the plane can be covered by the coins so that no two coins overlap~\cite{inaba2008puzzle}. Inaba also gave an elegant probabilistic proof of this result~\cite{inaba2008answer}, and asked for the largest number of points that can always be covered.

\begin{theorem}[Inaba]
  Any configuration of 10 points in the plane can be covered by pairwise non-overlapping coins.
\end{theorem}

The problem was later popularized by Winkler in \emph{Communications of the ACM} in 2010~\cite{winkler2010figures,winkler2010solutions}. Winkler showed that there exists a configuration of 60 points that cannot be covered by disjoint coins, using a suitable finite subset of the triangular lattice. More recently, the problem was presented by Sanderson in an online video~\cite{sanderson2026covering}. Throughout this paper, we model a coin as a closed unit disk.

Progress on the original finite problem has been made on both sides. On the lower-bound side, Aloupis, Hearn, Iwasawa, and Uehara showed that any configuration of 12 points can always be covered by disjoint unit disks~\cite{aloupis2012covering}. On the upper-bound side, Elser reduced the bound to 55 points by choosing a refined bounding box for a triangular-lattice construction~\cite{elser2011packing}. Okayama, Kiyomi, and Uehara further improved the bound to 53 points by considering configurations derived from the triangular, square, and hexagonal lattices~\cite{okayama2012covering}. Aloupis, Hearn, Iwasawa, and Uehara later reduced the bound to 50 points, again using the triangular lattice, and also proposed a computer-assisted heuristic construction of 45 points arranged on three concentric circles~\cite{aloupis2012covering}.

A related, but distinct, exact-cover variant was recently studied by Chun, Kipp, and Roch. In this variant, disks are allowed to overlap, but each input point must be covered exactly once. They proved that any 17-point set admits such an exact cover, and constructed a 657-point set for which no exact cover exists~\cite{chun2025exact}.

The lattice-based configurations used for the finite problem naturally suggest an infinite version. Given a lattice, can all of its points be covered by infinitely many pairwise disjoint unit disks? Let $d$ denote the minimum distance between two distinct lattice points; we refer to $d$ as the \emph{lattice spacing}. If $d$ is sufficiently large, then each lattice point can simply be covered by its own disk. Conversely, if the lattice is sufficiently dense, then a disk must cover multiple lattice points, and the non-overlap constraint becomes more restrictive.

Alm, Hommowun, Manary, and Schneider studied this infinite covering problem for the square lattice by introducing motif-based constructions that establish coverability for several intervals of the lattice spacing~\cite{alm2013motif}. Their motif framework provides a useful method for constructing periodic coverings of infinite lattices.

\begin{figure}[b]
  \centering
  \begin{tikzpicture}[scale=20]

  \coordinate (lowlim)    at (0.13,0);
  \coordinate (lowhex)    at (0.154700,0);  
  \coordinate (lowsq)     at (0.218780,0);  
  \coordinate (lowtri)    at (0.267949,0);  
  \coordinate (2sqrt37)   at (0.328798,0);
  \coordinate (1sqrt9)    at (0.333333,0);
  \coordinate (2sqrt31)   at (0.359211,0);
  \coordinate (1sqrt7)    at (0.377694,0);
  \coordinate (2sqrt27)   at (0.384900,0);
  \coordinate (3sqrt57)   at (0.397360,0);
  \coordinate (2sqrt21)   at (0.436436,0);
  \coordinate (2sqrt19)   at (0.458831,0);
  \coordinate (1sqrt4)    at (0.5,0);
  \coordinate (5sqrt91)   at (0.524142,0);
  \coordinate (6sqrt130)  at (0.526345,0);
  \coordinate (2sqrt13)   at (0.554700,0);
  \coordinate (2sqrt10)   at (0.632456,0);
  \coordinate (3sqrt21)   at (0.654654,0);
  \coordinate (4sqrt37)   at (0.657596,0);
  \coordinate (1sqrt2)    at (0.707107,0);
  \coordinate (5sqrt49)   at (0.714286,0);
  \coordinate (4sqrt31)   at (0.718421,0);
  \coordinate (4sqrt26)   at (0.784465,0);
  \coordinate (infinity)  at (0.83,0);

  \begin{scope}[gray!40]
    \draw ([yshift=-0.005cm]lowtri)  -- ([yshift=0.075cm]lowtri);
    \draw ([yshift=-0.005cm]2sqrt37) -- ([yshift=0.075cm]2sqrt37);
    \draw ([yshift=-0.005cm]2sqrt27) -- ([yshift=0.075cm]2sqrt27);
    \draw ([yshift=-0.005cm]2sqrt19) -- ([yshift=0.075cm]2sqrt19);
    \draw ([yshift=-0.005cm]2sqrt13) -- ([yshift=0.075cm]2sqrt13);
    \draw ([yshift=-0.005cm]4sqrt31) -- ([yshift=0.075cm]4sqrt31);
    \draw ([yshift=-0.005cm]4sqrt37) -- ([yshift=0.075cm]4sqrt37);
    \draw ([yshift=-0.005cm]lowhex)  -- ([yshift=0.050cm]lowhex);
    \draw ([yshift=-0.005cm]lowsq)   -- ([yshift=0.025cm]lowsq);
    \draw ([yshift=-0.005cm]1sqrt4)  -- ([yshift=0.025cm]1sqrt4);
    \draw ([yshift=-0.005cm]4sqrt26) -- ([yshift=0.025cm]4sqrt26);
    \draw[<->, gray] (0.05,0) -- (infinity);
  \end{scope}

  \begin{scope}[anchor=north east, font=\tiny]
    \node[rotate=30] at (lowhex)   {$r$ {\color{gray} (0.155)}};
    \node[rotate=30] at (lowsq)    {$\sqrt2\,r$ {\color{gray} (0.219)}};
    \node[rotate=30] at (lowtri)   {$\sqrt3\,r$ {\color{gray} (0.268)}};
    \node[rotate=30] at (2sqrt37)  {$2/\sqrt{37}$ {\color{gray} (0.329)}};
    \node[rotate=30] at (2sqrt27)  {$2/\sqrt{27}$ {\color{gray} (0.385)}};
    \node[rotate=30] at (2sqrt19)  {$2/\sqrt{19}$ {\color{gray} (0.459)}};
    \node[rotate=30] at (1sqrt4)   {$1/2$ {\color{gray} (0.500)}};
    \node[rotate=30] at (2sqrt13)  {$2/\sqrt{13}$ {\color{gray} (0.555)}};
    \node[rotate=30] at (4sqrt37)  {$4/\sqrt{37}$ {\color{gray} (0.658)}};
    \node[rotate=30] at (4sqrt31)  {$4/\sqrt{31}$ {\color{gray} (0.718)}};
    \node[rotate=30] at (4sqrt26)  {$4/\sqrt{26}$ {\color{gray} (0.784)}};
  \end{scope}

  \begin{scope}[thick, font=\scriptsize, transform canvas={yshift=1.5cm}]
    \draw (lowlim) node[anchor=east] {triangular};
    \draw[<-, red] (lowlim) -- (lowtri);
    \draw (2sqrt37) -- (1sqrt9);
    \draw (2sqrt31) -- (1sqrt7);
    \draw (2sqrt27) -- (3sqrt57);
    \draw (2sqrt19) -- (5sqrt91);
    \draw (2sqrt13) -- (3sqrt21);
    \draw (4sqrt37) -- (5sqrt49);
    \draw[->] (4sqrt31) -- (infinity);
  \end{scope}

  \begin{scope}[thick, font=\scriptsize, transform canvas={yshift=1.0cm}]
    \draw (lowlim) node[anchor=east] {honeycomb};
    \draw[<-, red] (lowlim) -- (lowhex);
    \draw (2sqrt37) -- (1sqrt7);
    \draw (2sqrt27) -- (3sqrt57);
    \draw (2sqrt21) -- (5sqrt91);
    \draw (2sqrt13) -- (3sqrt21);
    \draw[->] (4sqrt37) -- (infinity);
  \end{scope}

  \begin{scope}[thick, font=\scriptsize, transform canvas={yshift=0.5cm}]
    \draw (lowlim) node[anchor=east] {square};
    \draw[<-, red] (lowlim) -- (lowsq);
    \draw (1sqrt4) -- (6sqrt130);
    \draw (2sqrt13) -- (2sqrt10);
    \draw (4sqrt37) -- (1sqrt2);
    \draw[->] (4sqrt26) -- (infinity);
  \end{scope}

\end{tikzpicture}
  \caption{Coverable (black) and non-coverable (red) spacing ranges for the triangular lattice, honeycomb point set, and square lattice. Black intervals are certified by our constructions; red ranges were proved non-coverable by Okayama et~al.~\cite{okayama2012covering}. Here, $r=2/\sqrt3-1$.}
  \label{fig:overall-spacing}
\end{figure}
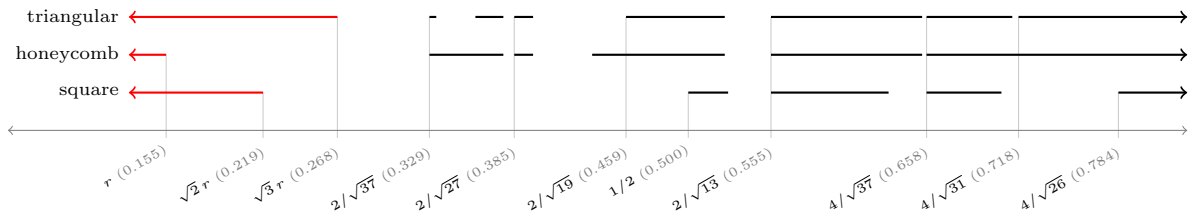

In this paper, we study the infinite lattice-covering problem for the three periodic point sets appearing in the constructions of Okayama et~al.: the triangular lattice, the square lattice, and the honeycomb point set, which they refer to as the hexagonal lattice. Our constructions establish several, generally disjoint, intervals of coverable lattice spacings, summarized in \cref{fig:overall-spacing}.

We first derive several elementary motif patterns in which each unit disk covers the same small number of lattice points. The case in which each disk covers exactly two lattice points is more subtle: a direct attempt at such a uniform construction leads to an obstruction. We overcome this obstruction by introducing a non-uniform periodic pattern in which different disks cover different numbers of lattice points; see \cref{fig:tri-cyclone}. The ideas underlying this construction also lead to simpler patterns and to a more general family of motifs, yielding several disjoint intervals of lattice spacings for which the triangular lattice is coverable. These constructions are discussed in detail in \cref{sec:tri}. We then briefly discuss the honeycomb point set in \cref{sec:hex}.

Finally, we revisit the square lattice in \cref{sec:square}. We identify an unintended overlap in one of the motif constructions of Alm et~al., which leaves a gap in the proof of their claimed coverability interval. We then give a modified motif that recovers part of this interval, and introduce a new motif construction that establishes an additional interval.

Throughout the paper, we illustrate each lattice together with its mesh graph for visual clarity. The lattice points are the vertices of the graph.

\section{Preliminaries}
\label{sec:prelim}

Throughout this paper, we identify the Euclidean plane with $\mathbb{R}^2$. When convenient, we also represent points using the complex plane.

We model a \emph{coin} as a closed unit disk. For $c\in\mathbb{R}^2$, let
\[
  D(c) = \{ p \in \mathbb{R}^2 : \|p-c\| \leq 1 \}
\]
denote the closed unit disk centered at $c$. A point $p$ is \emph{covered} by this disk if $p \in D(c)$.  Two unit disks are said to \emph{overlap} if their interiors intersect. Equivalently, disks centered at $c$ and $c'$ overlap iff
\[
  \|c-c'\| < 2.
\]
When $\|c-c'\|=2$, the two disks are tangent; we also say that they \emph{kiss}.

Since disks are closed and tangencies are allowed, a point may be covered by more than one disk, but only when it lies on the boundaries of the relevant disks. In such a situation, we assign the point to only one disk, which is said to be \emph{responsible} for that point.

A collection of unit disks is said to be in \emph{hexagonal packing} if its centers form a triangular lattice of spacing $2$. In this arrangement, each disk kisses six other disks.

\begin{definition}[Spacing]
  Let $\mathcal{X}$ be a discrete point set in $\mathbb{R}^2$. The \emph{spacing} of $\mathcal{X}$ is the minimum distance between two distinct points of $\mathcal{X}$. We denote this value by $d$.
\end{definition}

We consider three periodic point sets: the triangular lattice, the square lattice, and the honeycomb point set. The last of these is the point set referred to as the ``hexagonal lattice'' by Okayama et~al. We write $\mathcal{T}$, $\mathcal{S}$, and $\mathcal{H}$ for the corresponding unscaled point sets. For $d>0$, define
\[
  \mathcal{T}_d = d\mathcal{T}, \qquad
  \mathcal{S}_d = d\mathcal{S}, \qquad
  \mathcal{H}_d = d\mathcal{H},
\]
where, for a point set $\mathcal{X} \subseteq \mathbb{R}^2$,
\[
  d\mathcal{X} = \{ dx : x \in \mathcal{X} \}.
\]
Thus, $\mathcal{T}_d$, $\mathcal{S}_d$, and $\mathcal{H}_d$ have spacing $d$. We also use $\mathcal{L}$ to denote any one of $\mathcal{T}$, $\mathcal{S}$, and $\mathcal{H}$, and write $\mathcal{L}_d=d\mathcal{L}$.

For the triangular lattice, it is convenient to use Eisenstein-integer coordinates. Let
\[
  \omega = -\frac12 + \frac{\sqrt3}{2}i = e^{2\pi i/3}.
\]
The unscaled triangular lattice is
\[
  \mathcal{T} = \{ a+b\omega : a,b\in\mathbb{Z} \}.
\]
Likewise, the unscaled square lattice is represented by the Gaussian integers
\[
  \mathcal{S} = \{ a+bi : a,b\in\mathbb{Z} \}.
\]
The honeycomb point set, however, can be obtained from the triangular lattice by deleting one of the three congruence classes modulo $3$. Specifically, define
\[
  \mathcal{H} = \{ a+b\omega : a,b\in\mathbb{Z},\ a+b \not\equiv 0 \pmod{3} \}.
\]
This removes the congruence class $a+b\equiv 0 \pmod 3$ and leaves two interlaced triangular sublattices. The resulting point set has spacing $1$. In particular, $\mathcal{H}$ is not a lattice in the usual sense, but rather the union of two translates of a triangular lattice.

\begin{figure}[!t]
  \centering
  \begin{subfigure}[t]{0.32\linewidth}
    \centering
    \begin{tikzpicture}[scale=1]
  \clip (-2.2500,-2.2500) rectangle (2.2500,2.2500);
  \begin{scope}[lightgray]
    \draw (-5.0000,-4.3301) -- (5.0000,-4.3301);
    \draw (-5.0000,-3.4641) -- (5.0000,-3.4641);
    \draw (-5.0000,-2.5981) -- (5.0000,-2.5981);
    \draw (-5.0000,-1.7320) -- (5.0000,-1.7320);
    \draw (-5.0000,-0.86602) -- (5.0000,-0.86602);
    \draw (-5.0000,0) -- (5.0000,0);
    \draw (-5.0000,0.86602) -- (5.0000,0.86602);
    \draw (-5.0000,1.7320) -- (5.0000,1.7320);
    \draw (-5.0000,2.5981) -- (5.0000,2.5981);
    \draw (-5.0000,3.4641) -- (5.0000,3.4641);
    \draw (-5.0000,4.3301) -- (5.0000,4.3301);
    \draw (-9.5000,-4.3301) -- (-4.5000,4.3301);
    \draw (-8.5000,-4.3301) -- (-3.5000,4.3301);
    \draw (-7.5000,-4.3301) -- (-2.5000,4.3301);
    \draw (-6.5000,-4.3301) -- (-1.5000,4.3301);
    \draw (-5.5000,-4.3301) -- (-0.50000,4.3301);
    \draw (-4.5000,-4.3301) -- (0.50000,4.3301);
    \draw (-3.5000,-4.3301) -- (1.5000,4.3301);
    \draw (-2.5000,-4.3301) -- (2.5000,4.3301);
    \draw (-1.5000,-4.3301) -- (3.5000,4.3301);
    \draw (-0.50000,-4.3301) -- (4.5000,4.3301);
    \draw (0.50000,-4.3301) -- (5.5000,4.3301);
    \draw (1.5000,-4.3301) -- (6.5000,4.3301);
    \draw (2.5000,-4.3301) -- (7.5000,4.3301);
    \draw (3.5000,-4.3301) -- (8.5000,4.3301);
    \draw (4.5000,-4.3301) -- (9.5000,4.3301);
    \draw (-4.5000,-4.3301) -- (-9.5000,4.3301);
    \draw (-3.5000,-4.3301) -- (-8.5000,4.3301);
    \draw (-2.5000,-4.3301) -- (-7.5000,4.3301);
    \draw (-1.5000,-4.3301) -- (-6.5000,4.3301);
    \draw (-0.50000,-4.3301) -- (-5.5000,4.3301);
    \draw (0.50000,-4.3301) -- (-4.5000,4.3301);
    \draw (1.5000,-4.3301) -- (-3.5000,4.3301);
    \draw (2.5000,-4.3301) -- (-2.5000,4.3301);
    \draw (3.5000,-4.3301) -- (-1.5000,4.3301);
    \draw (4.5000,-4.3301) -- (-0.50000,4.3301);
    \draw (5.5000,-4.3301) -- (0.50000,4.3301);
    \draw (6.5000,-4.3301) -- (1.5000,4.3301);
    \draw (7.5000,-4.3301) -- (2.5000,4.3301);
    \draw (8.5000,-4.3301) -- (3.5000,4.3301);
    \draw (9.5000,-4.3301) -- (4.5000,4.3301);
  \end{scope}
  \begin{scope}[black]
    \fill (-2.5000,-2.5981) circle[radius=0.04];
    \fill (-1.5000,-2.5981) circle[radius=0.04];
    \fill (-0.50000,-2.5981) circle[radius=0.04];
    \fill (0.50000,-2.5981) circle[radius=0.04];
    \fill (1.5000,-2.5981) circle[radius=0.04];
    \fill (2.5000,-2.5981) circle[radius=0.04];
    \fill (-3.0000,-1.7320) circle[radius=0.04];
    \fill (-2.0000,-1.7320) circle[radius=0.04];
    \fill (-1.0000,-1.7320) circle[radius=0.04];
    \fill (0,-1.7320) circle[radius=0.04];
    \fill (1.0000,-1.7320) circle[radius=0.04];
    \fill (2.0000,-1.7320) circle[radius=0.04];
    \fill (3.0000,-1.7320) circle[radius=0.04];
    \fill (-2.5000,-0.86602) circle[radius=0.04];
    \fill (-1.5000,-0.86602) circle[radius=0.04];
    \fill (-0.50000,-0.86602) circle[radius=0.04];
    \fill (0.50000,-0.86602) circle[radius=0.04];
    \fill (1.5000,-0.86602) circle[radius=0.04];
    \fill (2.5000,-0.86602) circle[radius=0.04];
    \fill (-3.0000,0) circle[radius=0.04];
    \fill (-2.0000,0) circle[radius=0.04];
    \fill (-1.0000,0) circle[radius=0.04];
    \fill (0,0) circle[radius=0.04];
    \fill (1.0000,0) circle[radius=0.04];
    \fill (2.0000,0) circle[radius=0.04];
    \fill (3.0000,0) circle[radius=0.04];
    \fill (-2.5000,0.86602) circle[radius=0.04];
    \fill (-1.5000,0.86602) circle[radius=0.04];
    \fill (-0.50000,0.86602) circle[radius=0.04];
    \fill (0.50000,0.86602) circle[radius=0.04];
    \fill (1.5000,0.86602) circle[radius=0.04];
    \fill (2.5000,0.86602) circle[radius=0.04];
    \fill (-3.0000,1.7320) circle[radius=0.04];
    \fill (-2.0000,1.7320) circle[radius=0.04];
    \fill (-1.0000,1.7320) circle[radius=0.04];
    \fill (0,1.7320) circle[radius=0.04];
    \fill (1.0000,1.7320) circle[radius=0.04];
    \fill (2.0000,1.7320) circle[radius=0.04];
    \fill (3.0000,1.7320) circle[radius=0.04];
    \fill (-2.5000,2.5981) circle[radius=0.04];
    \fill (-1.5000,2.5981) circle[radius=0.04];
    \fill (-0.50000,2.5981) circle[radius=0.04];
    \fill (0.50000,2.5981) circle[radius=0.04];
    \fill (1.5000,2.5981) circle[radius=0.04];
    \fill (2.5000,2.5981) circle[radius=0.04];
  \end{scope}
  \begin{scope}[anchor=north, xshift=-0.6em, font=\small]
    \draw (0,0)              node {$0$};
    \draw (1.0000,0)         node {$1$};
    \draw (-0.50000,0.86602) node {$\omega$};
    \draw (1.5000,-0.86602)  node {$1{-}\omega$};
    \draw (2.0000,1.7320)    node {$3{+}2\omega$};
  \end{scope}
\end{tikzpicture}
    \caption{Triangular lattice}
  \end{subfigure}
  \hfill
  \begin{subfigure}[t]{0.32\linewidth}
    \centering
    \begin{tikzpicture}[scale=1]
  \clip (-2.2500,-2.2500) rectangle (2.2500,2.2500);
  \begin{scope}[lightgray]
    \draw (-5.0000,-5.0000) -- (5.0000,-5.0000);
    \draw (-5.0000,-4.0000) -- (5.0000,-4.0000);
    \draw (-5.0000,-3.0000) -- (5.0000,-3.0000);
    \draw (-5.0000,-2.0000) -- (5.0000,-2.0000);
    \draw (-5.0000,-1.0000) -- (5.0000,-1.0000);
    \draw (-5.0000,0) -- (5.0000,0);
    \draw (-5.0000,1.0000) -- (5.0000,1.0000);
    \draw (-5.0000,2.0000) -- (5.0000,2.0000);
    \draw (-5.0000,3.0000) -- (5.0000,3.0000);
    \draw (-5.0000,4.0000) -- (5.0000,4.0000);
    \draw (-5.0000,5.0000) -- (5.0000,5.0000);
    \draw (-5.0000,-5.0000) -- (-5.0000,5.0000);
    \draw (-4.0000,-5.0000) -- (-4.0000,5.0000);
    \draw (-3.0000,-5.0000) -- (-3.0000,5.0000);
    \draw (-2.0000,-5.0000) -- (-2.0000,5.0000);
    \draw (-1.0000,-5.0000) -- (-1.0000,5.0000);
    \draw (0,-5.0000) -- (0,5.0000);
    \draw (1.0000,-5.0000) -- (1.0000,5.0000);
    \draw (2.0000,-5.0000) -- (2.0000,5.0000);
    \draw (3.0000,-5.0000) -- (3.0000,5.0000);
    \draw (4.0000,-5.0000) -- (4.0000,5.0000);
    \draw (5.0000,-5.0000) -- (5.0000,5.0000);
  \end{scope}
  \begin{scope}[black]
    \fill (-3.0000,-3.0000) circle[radius=0.04];
    \fill (-2.0000,-3.0000) circle[radius=0.04];
    \fill (-1.0000,-3.0000) circle[radius=0.04];
    \fill (0,-3.0000) circle[radius=0.04];
    \fill (1.0000,-3.0000) circle[radius=0.04];
    \fill (2.0000,-3.0000) circle[radius=0.04];
    \fill (3.0000,-3.0000) circle[radius=0.04];
    \fill (-3.0000,-2.0000) circle[radius=0.04];
    \fill (-2.0000,-2.0000) circle[radius=0.04];
    \fill (-1.0000,-2.0000) circle[radius=0.04];
    \fill (0,-2.0000) circle[radius=0.04];
    \fill (1.0000,-2.0000) circle[radius=0.04];
    \fill (2.0000,-2.0000) circle[radius=0.04];
    \fill (3.0000,-2.0000) circle[radius=0.04];
    \fill (-3.0000,-1.0000) circle[radius=0.04];
    \fill (-2.0000,-1.0000) circle[radius=0.04];
    \fill (-1.0000,-1.0000) circle[radius=0.04];
    \fill (0,-1.0000) circle[radius=0.04];
    \fill (1.0000,-1.0000) circle[radius=0.04];
    \fill (2.0000,-1.0000) circle[radius=0.04];
    \fill (3.0000,-1.0000) circle[radius=0.04];
    \fill (-3.0000,0) circle[radius=0.04];
    \fill (-2.0000,0) circle[radius=0.04];
    \fill (-1.0000,0) circle[radius=0.04];
    \fill (0,0) circle[radius=0.04];
    \fill (1.0000,0) circle[radius=0.04];
    \fill (2.0000,0) circle[radius=0.04];
    \fill (3.0000,0) circle[radius=0.04];
    \fill (-3.0000,1.0000) circle[radius=0.04];
    \fill (-2.0000,1.0000) circle[radius=0.04];
    \fill (-1.0000,1.0000) circle[radius=0.04];
    \fill (0,1.0000) circle[radius=0.04];
    \fill (1.0000,1.0000) circle[radius=0.04];
    \fill (2.0000,1.0000) circle[radius=0.04];
    \fill (3.0000,1.0000) circle[radius=0.04];
    \fill (-3.0000,2.0000) circle[radius=0.04];
    \fill (-2.0000,2.0000) circle[radius=0.04];
    \fill (-1.0000,2.0000) circle[radius=0.04];
    \fill (0,2.0000) circle[radius=0.04];
    \fill (1.0000,2.0000) circle[radius=0.04];
    \fill (2.0000,2.0000) circle[radius=0.04];
    \fill (3.0000,2.0000) circle[radius=0.04];
    \fill (-3.0000,3.0000) circle[radius=0.04];
    \fill (-2.0000,3.0000) circle[radius=0.04];
    \fill (-1.0000,3.0000) circle[radius=0.04];
    \fill (0,3.0000) circle[radius=0.04];
    \fill (1.0000,3.0000) circle[radius=0.04];
    \fill (2.0000,3.0000) circle[radius=0.04];
    \fill (3.0000,3.0000) circle[radius=0.04];
  \end{scope}
  \begin{scope}[anchor=north, xshift=-0.4em, font=\small]
    \draw (0,0)  node {$0$};
    \draw (1,0)  node {$1$};
    \draw (0,1)  node {$i$};
    \draw (1,2)  node {$1{+}2i$};
    \draw (2,-1) node {$2{-}i$};
  \end{scope}
\end{tikzpicture}
    \caption{Square lattice}
  \end{subfigure}
  \hfill
  \begin{subfigure}[t]{0.32\linewidth}
    \centering
    \begin{tikzpicture}[scale=1]
  \clip (-2.2500,-2.2500) rectangle (2.2500,2.2500);
  \begin{scope}[lightgray]
    \draw (1.7320,2.0000) -- (0.86602,2.5000);
    \draw (-0.86602,0.50000) -- (-0.86602,-0.50000);
    \draw (0,-2.0000) -- (-0.86602,-2.5000);
    \draw (0,-2.0000) -- (0,-1.0000);
    \draw (0.86602,-0.50000) -- (0,-1.0000);
    \draw (0,2.0000) -- (0.86602,2.5000);
    \draw (2.5981,-2.5000) -- (1.7320,-2.0000);
    \draw (1.7320,1.0000) -- (1.7320,2.0000);
    \draw (0.86602,0.50000) -- (0,1.0000);
    \draw (1.7320,-2.0000) -- (0.86602,-2.5000);
    \draw (1.7320,-2.0000) -- (1.7320,-1.0000);
    \draw (2.5981,-0.50000) -- (1.7320,-1.0000);
    \draw (1.7320,1.0000) -- (0.86602,0.50000);
    \draw (-1.7320,1.0000) -- (-1.7320,2.0000);
    \draw (-1.7320,-1.0000) -- (-2.5981,-0.50000);
    \draw (-1.7320,2.0000) -- (-2.5981,2.5000);
    \draw (0,2.0000) -- (-0.86602,2.5000);
    \draw (-1.7320,2.0000) -- (-0.86602,2.5000);
    \draw (0,-2.0000) -- (0.86602,-2.5000);
    \draw (0.86602,-0.50000) -- (1.7320,-1.0000);
    \draw (0.86602,-0.50000) -- (0.86602,0.50000);
    \draw (-0.86602,-2.5000) -- (-1.7320,-2.0000);
    \draw (0,-1.0000) -- (-0.86602,-0.50000);
    \draw (2.5981,2.5000) -- (1.7320,2.0000);
    \draw (1.7320,1.0000) -- (2.5981,0.50000);
    \draw (0,2.0000) -- (0,1.0000);
    \draw (-0.86602,0.50000) -- (0,1.0000);
    \draw (-1.7320,1.0000) -- (-0.86602,0.50000);
    \draw (-1.7320,1.0000) -- (-2.5981,0.50000);
    \draw (-2.5981,-2.5000) -- (-1.7320,-2.0000);
    \draw (-1.7320,-1.0000) -- (-1.7320,-2.0000);
    \draw (-1.7320,-1.0000) -- (-0.86602,-0.50000);
  \end{scope}
  \begin{scope}[black]
    \fill (-2.5981,-2.5000) circle[radius=0.04];
    \fill (-0.86602,-2.5000) circle[radius=0.04];
    \fill (0.86602,-2.5000) circle[radius=0.04];
    \fill (2.5981,-2.5000) circle[radius=0.04];
    \fill (-1.7320,-1.0000) circle[radius=0.04];
    \fill (0,-1.0000) circle[radius=0.04];
    \fill (1.7320,-1.0000) circle[radius=0.04];
    \fill (-2.5981,0.50000) circle[radius=0.04];
    \fill (-0.86602,0.50000) circle[radius=0.04];
    \fill (0.86602,0.50000) circle[radius=0.04];
    \fill (2.5981,0.50000) circle[radius=0.04];
    \fill (-1.7320,2.0000) circle[radius=0.04];
    \fill (0,2.0000) circle[radius=0.04];
    \fill (1.7320,2.0000) circle[radius=0.04];
  \end{scope}
  \begin{scope}[black]
    \fill (-1.7320,-2.0000) circle[radius=0.04];
    \fill (0,-2.0000) circle[radius=0.04];
    \fill (1.7320,-2.0000) circle[radius=0.04];
    \fill (-2.5981,-0.50000) circle[radius=0.04];
    \fill (-0.86602,-0.50000) circle[radius=0.04];
    \fill (0.86602,-0.50000) circle[radius=0.04];
    \fill (2.5981,-0.50000) circle[radius=0.04];
    \fill (-1.7320,1.0000) circle[radius=0.04];
    \fill (0,1.0000) circle[radius=0.04];
    \fill (1.7320,1.0000) circle[radius=0.04];
    \fill (-2.5981,2.5000) circle[radius=0.04];
    \fill (-0.86602,2.5000) circle[radius=0.04];
    \fill (0.86602,2.5000) circle[radius=0.04];
    \fill (2.5981,2.5000) circle[radius=0.04];
  \end{scope}
  \begin{scope}[anchor=north, xshift=-0.6em, font=\small]
    \draw (0.86602,0.5)  node {$1$};
    \draw (-0.86602,0.5) node {$\omega$};
    \draw (1.7320,-1)    node {$-2\omega$};
    \draw (-1.7320,-1)   node {$-2$};
    \draw (0,2)          node {$2{+}2\omega$};
  \end{scope}
\end{tikzpicture}
    \caption{Honeycomb point set}
  \end{subfigure}
  \caption{Coordinate representations using complex numbers, independent of the spacing $d$.}
  \label{fig:lattices}
\end{figure}
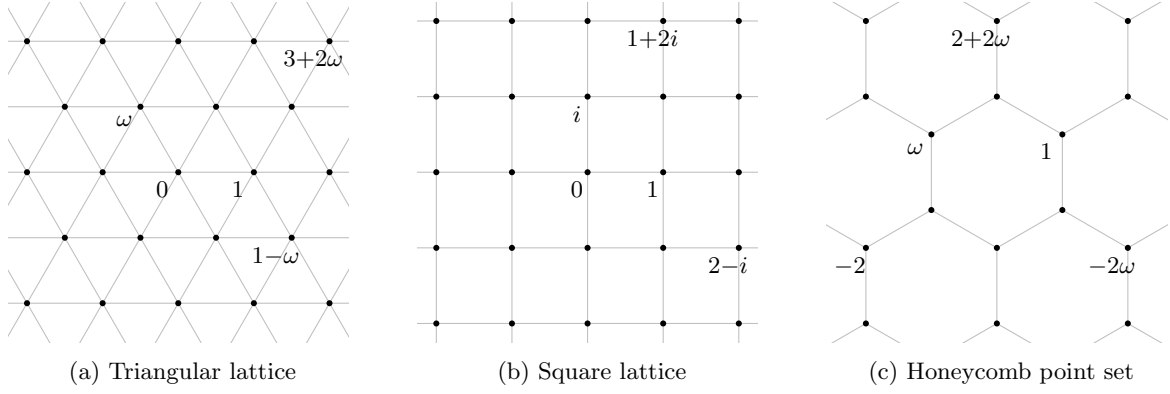

Whenever convenient, we represent points in these sets by complex numbers and convert them to Cartesian coordinates. For the triangular lattice $\mathcal{T}_d$, an Eisenstein-integer is mapped as follows:
\[
  a+b\omega
  \quad\longmapsto\quad
  \left(
    \frac{(2a-b)d}{2}
    ,
    \frac{\sqrt3\,bd}{2}
  \right).
\]
The same convention applies to Gaussian-integer coordinates for the square lattice $\mathcal{S}_d$. \Cref{fig:lattices} illustrates these coordinate representations.

We now describe the motif framework used throughout the paper. The framework is adapted from the motif constructions of Alm et~al.~\cite{alm2013motif}. In our formulation, motif copies are generated by translations only, but a motif pattern may consist of several motif families. This differs slightly from the framework of Alm et~al., where copies of a motif may be obtained by more general isometries preserving the underlying point set.

\begin{definition}[Motif]
  A \emph{motif} over $\mathcal{L}_d$ is a finite subset of $\mathcal{L}_d$.
\end{definition}

Intuitively, a motif represents the set of lattice points assigned to a single disk. Not every motif is necessarily realizable by a unit disk, and even when a motif is realizable, the realizing disk need not be unique.

\begin{definition}[Realization of a motif]
A unit disk $D(c)$ is said to \emph{realize} a motif $M\subseteq \mathcal{L}_d$ if
  \[
    \operatorname{int} D(c) \cap \mathcal{L}_d
    \:\subseteq\:
    M 
    \:\subseteq\:
    D(c)\cap \mathcal{L}_d.
  \]
\end{definition}

Equivalently, all lattice points in $M$ are covered by the disk, and every lattice point of $\mathcal{L}_d$ covered by the interior of the disk belongs to $M$. Thus, lattice points not belonging to $M$ may still lie on the boundary of $D(c)$. This convention is needed because disks are closed and tangent disks are allowed; a lattice point lying at a tangency point may therefore be covered by two non-overlapping disks.

When we say that a disk is \emph{responsible for} $k$ lattice points, we mean that its associated motif has cardinality $k$. The disk may contain additional lattice points on its boundary, but these boundary incidences are not counted as part of the motif unless the points are assigned to that disk.

A natural candidate for realizing a motif is obtained by placing the disk center at a Chebyshev center of the motif, that is, at the center of a minimum-radius disk containing the motif. This candidate is useful in several of our constructions, although it does not always yield a valid realization.

\begin{definition}[Translation lattice]
  A \emph{translation lattice} for $\mathcal{L}_d$ is a lattice $\Lambda\subseteq\mathbb{R}^2$ such that
  \[
    \forall v \in \Lambda,
    \quad
    \mathcal{L}_d + v = \mathcal{L}_d.
  \]
\end{definition}

For the triangular and square lattices, a translation lattice may be chosen as a sublattice of the point set itself. For the honeycomb point set, however, it is a lattice of translation symmetries and is not a subset of $\mathcal{H}_d$.

\begin{definition}[Motif pattern]
  Let $M\subseteq\mathcal{L}_d$ be a motif, and let $\Lambda$ be a translation lattice for $\mathcal{L}_d$. The \emph{motif pattern} generated by $M$ and $\Lambda$ is
  \[
    \mathcal{P}(M,\Lambda) = \{ M + v : v \in \Lambda \},
  \]
  provided that the motif copies are pairwise disjoint.

  More generally, an \emph{$n$-family motif pattern} is a union
  \[
    \mathcal{P} =
    \mathcal{P}^{(1)}
    \cup
    \mathcal{P}^{(2)}
    \cup \ldots \cup
    \mathcal{P}^{(n)},
  \]
  where all motif copies appearing in the union are pairwise disjoint.
\end{definition}

A motif pattern \emph{covers} $\mathcal{L}_d$ if the union of all motif copies in the pattern is $\mathcal{L}_d$. A \emph{realization} of a motif pattern consists of choosing a realizing disk for every motif copy. The realization is \emph{valid} if no pair of the chosen disks overlap. Therefore, a valid realization of a motif pattern that covers $\mathcal{L}_d$ gives a covering of all points of $\mathcal{L}_d$ by pairwise non-overlapping unit disks.

All motif patterns considered in this paper are periodic. We use the term \emph{tile} to refer to the finite data in one fundamental region: the motif copies in that region, together with their realizing disks. Repeating the tile by the period translations generates the full infinite motif pattern.

\section{Triangular Lattice}
\label{sec:tri}

In this section, we give periodic constructions for covering the triangular lattice by pairwise non-overlapping unit disks. The main result of this section is the following.

\begin{theorem}\label{thm:tri-spacing}
  The triangular lattice $\mathcal{T}_d$ can be covered by pairwise non-overlapping unit disks whenever
  \[
    d \in
    \left[ \frac2{\sqrt{37}} , \frac13 \right]
    \cup
    \left[ \frac2{\sqrt{31}} , \frac1{\sqrt7} \right]
    \cup
    \left[ \frac2{\sqrt{27}} , \frac3{\sqrt{57}} \right]
    \cup
    \left[ \frac2{\sqrt{19}} , \frac5{\sqrt{91}} \right]
    \cup
    \left[ \frac2{\sqrt{13}} , \frac3{\sqrt{21}} \right]
    \cup
    \left[ \frac4{\sqrt{37}} , \frac57 \right]
    \cup
    \left[ \frac4{\sqrt{31}} , \infty \right).
  \]
\end{theorem}

The proof is constructive. Every spacing in \cref{thm:tri-spacing} is certified by one of the periodic motif pattern together with a valid realization. Since all motifs in this section are specified by fixed relative positions in $\mathcal{T}_d$, all relevant distances scale linearly with $d$. Thus, for each motif pattern, validity reduces to checking two types of inequalities: assigned lattice points must lie in their realizing disks, and the centers of realizing disks must be at distance at least $2$ from one another. The endpoints of the intervals below are obtained from these active constraints.

For reference, \cref{tab:tri-patterns} summarizes the motif patterns used in this section. Taking the union of the listed intervals gives \cref{thm:tri-spacing}.

\begin{table}[!b]
  \centering
  \small
  \begin{tabular}{lcccccc}
    \toprule
    \multirow{2}{*}{Pattern}
      & \multirow{2}{*}{Centers}
      & \multirow{2}{*}{Responsibilities}
      & \multicolumn{2}{c}{$d_{\min}$} & \multicolumn{2}{c}{$d_{\max}$} \\
    \cmidrule(lr){4-5}\cmidrule(lr){6-7}
      & & & Exact & Approx. & Exact & Approx. \\
    \midrule
    Singleton
      & vertex
      & $1$
      & $2$
      & $2.000$
      & $\infty$
      & - \\

    Triangle
      & face
      & $3$
      & $2/\sqrt3$
      & $1.155$
      & $\sqrt3$
      & $1.732$ \\

    Rhombus
      & edge
      & $4$
      & $1$
      & $1.000$
      & $2/\sqrt3$
      & $1.155$ \\

    Hexagon
      & vertex
      & $7$
      & $2/\sqrt7$
      & $0.756$
      & $1$
      & $1.000$ \\

    Alt.\ pair
      & edges
      & $2+2$
      & $4/\sqrt7$
      & $1.512$
      & $2$
      & $2.000$ \\

    Four-family
      & vertex and edges
      & $1+2+2+2$
      & $4/\sqrt7$
      & $1.512$
      & $2$
      & $2.000$ \\

    Single-family
      & vertex
      & $13$
      & $2/\sqrt{13}$
      & $0.555$
      & $1/\sqrt3$
      & $0.577$ \\

    Single-family
      & vertex
      & $19$
      & $2/\sqrt{19}$
      & $0.459$
      & $1/2$
      & $0.500$ \\

    Single-family
      & vertex
      & $31$
      & $2/\sqrt{31}$
      & $0.359$
      & $1/\sqrt7$
      & $0.378$ \\

    Single-family
      & vertex
      & $37$
      & $2/\sqrt{37}$
      & $0.329$
      & $1/3$
      & $0.333$ \\

    Single-family
      & face
      & $12$
      & $1/\sqrt3$
      & $0.577$
      & $3/\sqrt{21}$
      & $0.655$ \\

    Single-family
      & face
      & $27$
      & $2/\sqrt{27}$
      & $0.385$
      & $3/\sqrt{57}$
      & $0.397$ \\

    Single-family
      & Chebyshev
      & $9$
      & $2/3$
      & $0.667$
      & $5/7$
      & $0.714$ \\

    Single-family
      & Chebyshev
      & $16$
      & $1/2$
      & $0.500$
      & $5/\sqrt{91}$
      & $0.524$ \\

    Four-family
      & vertex and edges
      & $7+8+8+8$
      & $4/\sqrt{31}$
      & $0.718$
      & $2/\sqrt7$
      & $0.756$ \\

    Four-family
      & vertex and edges
      & $7+10+10+10$
      & $4/\sqrt{37}$
      & $0.658$
      & $2/3$
      & $0.667$ \\
    \bottomrule
  \end{tabular}
  \caption{Periodic motif patterns for the triangular lattice.}
  \label{tab:tri-patterns}
\end{table}

\subsection{Primitive motif patterns}
\label{sec:tri-primitive}

We begin with four elementary single-family motif patterns. The simplest case assigns one disk to each lattice point. This gives a valid covering precisely when the disks centered at adjacent lattice points do not overlap, namely when $d\ge 2$. At the endpoint $d=2$, the disks form the usual hexagonal packing.

\begin{figure}[!h]
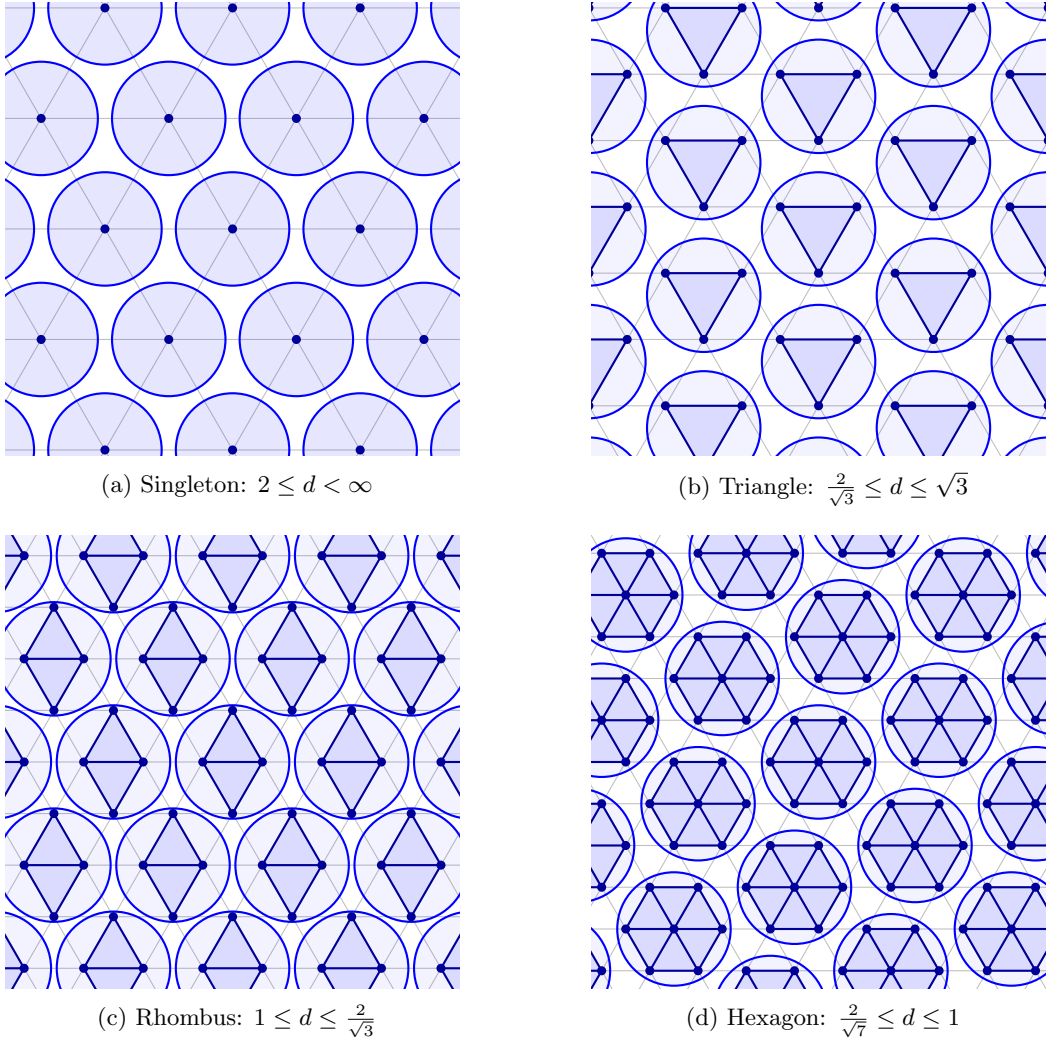

  \centering
  \begin{subfigure}[t]{0.45\linewidth}
    \centering
    \begin{tikzpicture}[scale=0.75]
  \clip (-4.0000,-4.0000) rectangle (4.0000,4.0000);
  \begin{scope}[lightgray]
    \draw (-9.0000,-9.7428) -- (9.0000,-9.7428);
    \draw (-9.0000,-7.7942) -- (9.0000,-7.7942);
    \draw (-9.0000,-5.8457) -- (9.0000,-5.8457);
    \draw (-9.0000,-3.8971) -- (9.0000,-3.8971);
    \draw (-9.0000,-1.9486) -- (9.0000,-1.9486);
    \draw (-9.0000,0) -- (9.0000,0);
    \draw (-9.0000,1.9486) -- (9.0000,1.9486);
    \draw (-9.0000,3.8971) -- (9.0000,3.8971);
    \draw (-9.0000,5.8457) -- (9.0000,5.8457);
    \draw (-9.0000,7.7942) -- (9.0000,7.7942);
    \draw (-9.0000,9.7428) -- (9.0000,9.7428);
    \draw (-19.125,-9.7428) -- (-7.8750,9.7428);
    \draw (-16.875,-9.7428) -- (-5.6250,9.7428);
    \draw (-14.625,-9.7428) -- (-3.3750,9.7428);
    \draw (-12.375,-9.7428) -- (-1.1250,9.7428);
    \draw (-10.125,-9.7428) -- (1.1250,9.7428);
    \draw (-7.8750,-9.7428) -- (3.3750,9.7428);
    \draw (-5.6250,-9.7428) -- (5.6250,9.7428);
    \draw (-3.3750,-9.7428) -- (7.8750,9.7428);
    \draw (-1.1250,-9.7428) -- (10.125,9.7428);
    \draw (1.1250,-9.7428) -- (12.375,9.7428);
    \draw (3.3750,-9.7428) -- (14.625,9.7428);
    \draw (5.6250,-9.7428) -- (16.875,9.7428);
    \draw (7.8750,-9.7428) -- (19.125,9.7428);
    \draw (-7.8750,-9.7428) -- (-19.125,9.7428);
    \draw (-5.6250,-9.7428) -- (-16.875,9.7428);
    \draw (-3.3750,-9.7428) -- (-14.625,9.7428);
    \draw (-1.1250,-9.7428) -- (-12.375,9.7428);
    \draw (1.1250,-9.7428) -- (-10.125,9.7428);
    \draw (3.3750,-9.7428) -- (-7.8750,9.7428);
    \draw (5.6250,-9.7428) -- (-5.6250,9.7428);
    \draw (7.8750,-9.7428) -- (-3.3750,9.7428);
    \draw (10.125,-9.7428) -- (-1.1250,9.7428);
    \draw (12.375,-9.7428) -- (1.1250,9.7428);
    \draw (14.625,-9.7428) -- (3.3750,9.7428);
    \draw (16.875,-9.7428) -- (5.6250,9.7428);
    \draw (19.125,-9.7428) -- (7.8750,9.7428);
  \end{scope}
  \begin{scope}[blue, thick, fill opacity=0.1]
    \filldraw (-4.5000,-3.8971) circle[radius=1];
    \filldraw (-2.2500,-3.8971) circle[radius=1];
    \filldraw (0,-3.8971) circle[radius=1];
    \filldraw (2.2500,-3.8971) circle[radius=1];
    \filldraw (4.5000,-3.8971) circle[radius=1];
    \filldraw (-3.3750,-1.9486) circle[radius=1];
    \filldraw (-1.1250,-1.9486) circle[radius=1];
    \filldraw (1.1250,-1.9486) circle[radius=1];
    \filldraw (3.3750,-1.9486) circle[radius=1];
    \filldraw (-4.5000,0) circle[radius=1];
    \filldraw (-2.2500,0) circle[radius=1];
    \filldraw (0,0) circle[radius=1];
    \filldraw (2.2500,0) circle[radius=1];
    \filldraw (4.5000,0) circle[radius=1];
    \filldraw (-3.3750,1.9486) circle[radius=1];
    \filldraw (-1.1250,1.9486) circle[radius=1];
    \filldraw (1.1250,1.9486) circle[radius=1];
    \filldraw (3.3750,1.9486) circle[radius=1];
    \filldraw (-4.5000,3.8971) circle[radius=1];
    \filldraw (-2.2500,3.8971) circle[radius=1];
    \filldraw (0,3.8971) circle[radius=1];
    \filldraw (2.2500,3.8971) circle[radius=1];
    \filldraw (4.5000,3.8971) circle[radius=1];
  \end{scope}
  \begin{scope}[blue, fill opacity=0.1]
  \end{scope}
  \begin{scope}[blue!60!black, thick]
    \fill (-4.5000,-3.8971) circle[radius=0.080000];
    \fill (-2.2500,-3.8971) circle[radius=0.080000];
    \fill (0,-3.8971) circle[radius=0.080000];
    \fill (2.2500,-3.8971) circle[radius=0.080000];
    \fill (4.5000,-3.8971) circle[radius=0.080000];
    \fill (-3.3750,-1.9486) circle[radius=0.080000];
    \fill (-1.1250,-1.9486) circle[radius=0.080000];
    \fill (1.1250,-1.9486) circle[radius=0.080000];
    \fill (3.3750,-1.9486) circle[radius=0.080000];
    \fill (-4.5000,0) circle[radius=0.080000];
    \fill (-2.2500,0) circle[radius=0.080000];
    \fill (0,0) circle[radius=0.080000];
    \fill (2.2500,0) circle[radius=0.080000];
    \fill (4.5000,0) circle[radius=0.080000];
    \fill (-3.3750,1.9486) circle[radius=0.080000];
    \fill (-1.1250,1.9486) circle[radius=0.080000];
    \fill (1.1250,1.9486) circle[radius=0.080000];
    \fill (3.3750,1.9486) circle[radius=0.080000];
    \fill (-4.5000,3.8971) circle[radius=0.080000];
    \fill (-2.2500,3.8971) circle[radius=0.080000];
    \fill (0,3.8971) circle[radius=0.080000];
    \fill (2.2500,3.8971) circle[radius=0.080000];
    \fill (4.5000,3.8971) circle[radius=0.080000];
  \end{scope}
\end{tikzpicture}
    \caption{Singleton: $2 \le d < \infty$}
    \label{fig:tri-uniform1}
  \end{subfigure}
  \hspace{1em}
  \begin{subfigure}[t]{0.45\linewidth}
    \centering
    \input{img/tri/uniform3.tikz}
    \caption{Triangle: $\frac{2}{\sqrt3} \le d \le \sqrt3$}
    \label{fig:tri-uniform3}
  \end{subfigure}
  \par\vspace{1em}
  \begin{subfigure}[t]{0.45\linewidth}
    \centering
    \input{img/tri/uniform4.tikz}
    \caption{Rhombus: $1 \le d \le \frac{2}{\sqrt3}$}
    \label{fig:tri-uniform4}
  \end{subfigure}
  \hspace{1em}
  \begin{subfigure}[t]{0.45\linewidth}
    \centering
    \input{img/tri/uniform7.tikz}
    \caption{Hexagon: $\frac{2}{\sqrt7} \le d \le 1$}
    \label{fig:tri-uniform7}
  \end{subfigure}
  \caption{Primitive single-family motif patterns with realizations for the triangular lattice.}
  \label{fig:tri-primitive}
\end{figure}

The same hexagonal-packing structure can be used for finer triangular lattices by assigning several lattice points to each disk. In this way we obtain uniform motif patterns in which each disk is responsible for $3$, $4$, or $7$ lattice points; see \cref{fig:tri-primitive}. In each case, the lower endpoint is the value of $d$ at which neighboring realizing disks kiss, while the upper endpoint is the value of $d$ at which the farthest assigned lattice point lies on the boundary of its realizing disk.

The four primitive patterns cover all spacings
\[
  d \in \left[\frac{2}{\sqrt7},\sqrt3\right] \cup [2,\infty).
\]
They therefore leave a gap for $\sqrt3 < d < 2$. In this range, a unit disk cannot cover three mutually adjacent lattice points, but assigning one disk to each lattice point would force adjacent disks to overlap. The natural next possibility is to assign two lattice points to each disk.

\begin{figure}[t]
  \centering
  \input{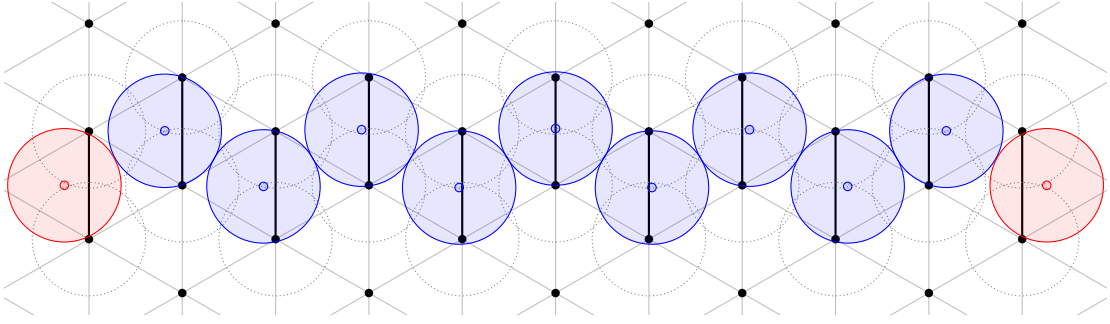}
  \caption{Example with $d=1.9$ illustrating the obstruction to realizing the two-point motif.}
  \label{fig:tri-failed}
\end{figure}

A uniform two-point construction is more delicate than the primitive patterns. The most direct attempt, in which each motif is responsible for a single edge of the triangular mesh, runs into the obstruction shown in \cref{fig:tri-failed}. Moving the realizing disks parallel to the edge does not remove all overlaps, while shifting the centers perpendicular to the edge causes disks in neighboring rows to miss their assigned lattice points.

The gap can nevertheless be covered by periodic multi-family patterns. One such construction is the ``cyclone'' pattern shown in \cref{fig:tri-cyclone}. This pattern motivated the use of non-uniform periodic motifs, but its description is somewhat lengthy; we give its coordinates and verification in \cref{app:tri-cyclone}.

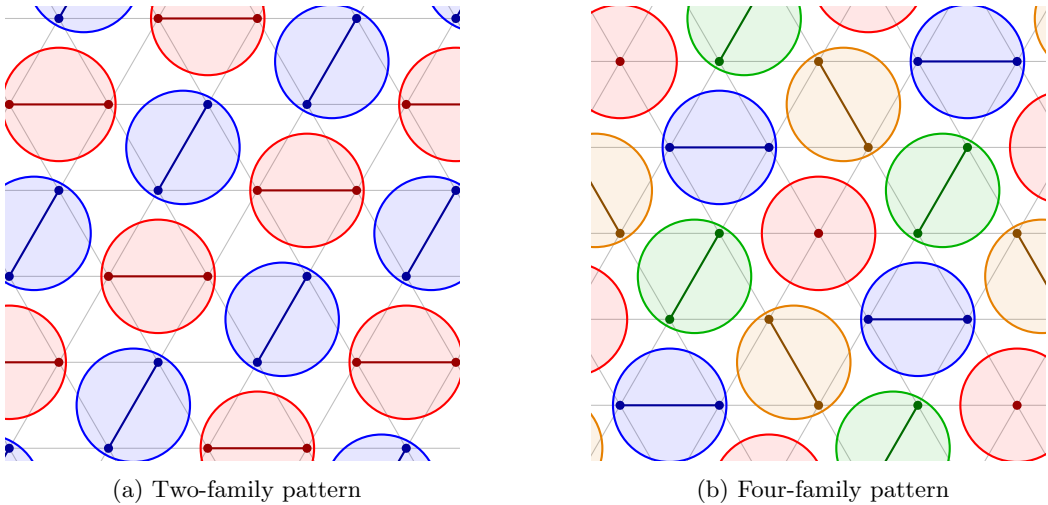
\begin{figure}[!b]
  \centering
  \begin{subfigure}[t]{0.45\linewidth}
    \centering
    \begin{tikzpicture}[scale=0.75]
  \clip (-1.8125,-3.2422) rectangle (6.1875,4.7578);
  \begin{scope}[lightgray]
    \draw (-10.500,-9.0933) -- (10.500,-9.0933);
    \draw (-10.500,-7.5777) -- (10.500,-7.5777);
    \draw (-10.500,-6.0622) -- (10.500,-6.0622);
    \draw (-10.500,-4.5466) -- (10.500,-4.5466);
    \draw (-10.500,-3.0311) -- (10.500,-3.0311);
    \draw (-10.500,-1.5155) -- (10.500,-1.5155);
    \draw (-10.500,0) -- (10.500,0);
    \draw (-10.500,1.5155) -- (10.500,1.5155);
    \draw (-10.500,3.0311) -- (10.500,3.0311);
    \draw (-10.500,4.5466) -- (10.500,4.5466);
    \draw (-10.500,6.0622) -- (10.500,6.0622);
    \draw (-10.500,7.5777) -- (10.500,7.5777);
    \draw (-10.500,9.0933) -- (10.500,9.0933);
    \draw (-21.000,-9.0933) -- (-10.500,9.0933);
    \draw (-19.250,-9.0933) -- (-8.7500,9.0933);
    \draw (-17.500,-9.0933) -- (-7.0000,9.0933);
    \draw (-15.750,-9.0933) -- (-5.2500,9.0933);
    \draw (-14.000,-9.0933) -- (-3.5000,9.0933);
    \draw (-12.250,-9.0933) -- (-1.7500,9.0933);
    \draw (-10.500,-9.0933) -- (0,9.0933);
    \draw (-8.7500,-9.0933) -- (1.7500,9.0933);
    \draw (-7.0000,-9.0933) -- (3.5000,9.0933);
    \draw (-5.2500,-9.0933) -- (5.2500,9.0933);
    \draw (-3.5000,-9.0933) -- (7.0000,9.0933);
    \draw (-1.7500,-9.0933) -- (8.7500,9.0933);
    \draw (0,-9.0933) -- (10.500,9.0933);
    \draw (1.7500,-9.0933) -- (12.250,9.0933);
    \draw (3.5000,-9.0933) -- (14.000,9.0933);
    \draw (5.2500,-9.0933) -- (15.750,9.0933);
    \draw (7.0000,-9.0933) -- (17.500,9.0933);
    \draw (8.7500,-9.0933) -- (19.250,9.0933);
    \draw (10.500,-9.0933) -- (21.000,9.0933);
    \draw (-10.500,-9.0933) -- (-21.000,9.0933);
    \draw (-8.7500,-9.0933) -- (-19.250,9.0933);
    \draw (-7.0000,-9.0933) -- (-17.500,9.0933);
    \draw (-5.2500,-9.0933) -- (-15.750,9.0933);
    \draw (-3.5000,-9.0933) -- (-14.000,9.0933);
    \draw (-1.7500,-9.0933) -- (-12.250,9.0933);
    \draw (0,-9.0933) -- (-10.500,9.0933);
    \draw (1.7500,-9.0933) -- (-8.7500,9.0933);
    \draw (3.5000,-9.0933) -- (-7.0000,9.0933);
    \draw (5.2500,-9.0933) -- (-5.2500,9.0933);
    \draw (7.0000,-9.0933) -- (-3.5000,9.0933);
    \draw (8.7500,-9.0933) -- (-1.7500,9.0933);
    \draw (10.500,-9.0933) -- (0,9.0933);
    \draw (12.250,-9.0933) -- (1.7500,9.0933);
    \draw (14.000,-9.0933) -- (3.5000,9.0933);
    \draw (15.750,-9.0933) -- (5.2500,9.0933);
    \draw (17.500,-9.0933) -- (7.0000,9.0933);
    \draw (19.250,-9.0933) -- (8.7500,9.0933);
    \draw (21.000,-9.0933) -- (10.500,9.0933);
  \end{scope}
  \begin{scope}[red, thick, fill opacity=0.1]
    \filldraw (-1.7500,-1.5155) circle[radius=1];
    \filldraw (-0.87500,3.0311) circle[radius=1];
    \filldraw (0.87500,0) circle[radius=1];
    \filldraw (2.6250,-3.0311) circle[radius=1];
    \filldraw (1.7500,4.5466) circle[radius=1];
    \filldraw (3.5000,1.5155) circle[radius=1];
    \filldraw (5.2500,-1.5155) circle[radius=1];
    \filldraw (6.1250,3.0311) circle[radius=1];
  \end{scope}
  \begin{scope}[blue, thick, fill opacity=0.1]
    \filldraw (-2.1875,-3.7889) circle[radius=1];
    \filldraw (-1.3125,0.75777) circle[radius=1];
    \filldraw (0.43750,-2.2733) circle[radius=1];
    \filldraw (-0.43750,5.3044) circle[radius=1];
    \filldraw (1.3125,2.2733) circle[radius=1];
    \filldraw (3.0625,-0.75777) circle[radius=1];
    \filldraw (4.8125,-3.7889) circle[radius=1];
    \filldraw (3.9375,3.7889) circle[radius=1];
    \filldraw (5.6875,0.75777) circle[radius=1];
    \filldraw (6.5625,5.3044) circle[radius=1];
  \end{scope}
  \begin{scope}[red, fill opacity=0.1]
  \end{scope}
  \begin{scope}[red!60!black, thick]
    \draw (-2.6250,-1.5155) -- (-0.87500,-1.5155);
    \fill (-2.6250,-1.5155) circle[radius=0.080000];
    \fill (-0.87500,-1.5155) circle[radius=0.080000];
    \draw (0,3.0311) -- (-1.7500,3.0311);
    \fill (0,3.0311) circle[radius=0.080000];
    \fill (-1.7500,3.0311) circle[radius=0.080000];
    \draw (0,0) -- (1.7500,0);
    \fill (0,0) circle[radius=0.080000];
    \fill (1.7500,0) circle[radius=0.080000];
    \draw (1.7500,-3.0311) -- (3.5000,-3.0311);
    \fill (1.7500,-3.0311) circle[radius=0.080000];
    \fill (3.5000,-3.0311) circle[radius=0.080000];
    \draw (0.87500,4.5466) -- (2.6250,4.5466);
    \fill (0.87500,4.5466) circle[radius=0.080000];
    \fill (2.6250,4.5466) circle[radius=0.080000];
    \draw (2.6250,1.5155) -- (4.3750,1.5155);
    \fill (2.6250,1.5155) circle[radius=0.080000];
    \fill (4.3750,1.5155) circle[radius=0.080000];
    \draw (4.3750,-1.5155) -- (6.1250,-1.5155);
    \fill (4.3750,-1.5155) circle[radius=0.080000];
    \fill (6.1250,-1.5155) circle[radius=0.080000];
    \draw (5.2500,3.0311) -- (7.0000,3.0311);
    \fill (5.2500,3.0311) circle[radius=0.080000];
    \fill (7.0000,3.0311) circle[radius=0.080000];
  \end{scope}
  \begin{scope}[blue, fill opacity=0.1]
  \end{scope}
  \begin{scope}[blue!60!black, thick]
    \draw (-1.7500,-3.0311) -- (-2.6250,-4.5466);
    \fill (-1.7500,-3.0311) circle[radius=0.080000];
    \fill (-2.6250,-4.5466) circle[radius=0.080000];
    \draw (-1.7500,0) -- (-0.87500,1.5155);
    \fill (-1.7500,0) circle[radius=0.080000];
    \fill (-0.87500,1.5155) circle[radius=0.080000];
    \draw (0.87500,-1.5155) -- (0,-3.0311);
    \fill (0.87500,-1.5155) circle[radius=0.080000];
    \fill (0,-3.0311) circle[radius=0.080000];
    \draw (-0.87500,4.5466) -- (0,6.0622);
    \fill (-0.87500,4.5466) circle[radius=0.080000];
    \fill (0,6.0622) circle[radius=0.080000];
    \draw (0.87500,1.5155) -- (1.7500,3.0311);
    \fill (0.87500,1.5155) circle[radius=0.080000];
    \fill (1.7500,3.0311) circle[radius=0.080000];
    \draw (2.6250,-1.5155) -- (3.5000,0);
    \fill (2.6250,-1.5155) circle[radius=0.080000];
    \fill (3.5000,0) circle[radius=0.080000];
    \draw (5.2500,-3.0311) -- (4.3750,-4.5466);
    \fill (5.2500,-3.0311) circle[radius=0.080000];
    \fill (4.3750,-4.5466) circle[radius=0.080000];
    \draw (3.5000,3.0311) -- (4.3750,4.5466);
    \fill (3.5000,3.0311) circle[radius=0.080000];
    \fill (4.3750,4.5466) circle[radius=0.080000];
    \draw (5.2500,0) -- (6.1250,1.5155);
    \fill (5.2500,0) circle[radius=0.080000];
    \fill (6.1250,1.5155) circle[radius=0.080000];
    \draw (6.1250,4.5466) -- (7.0000,6.0622);
    \fill (6.1250,4.5466) circle[radius=0.080000];
    \fill (7.0000,6.0622) circle[radius=0.080000];
  \end{scope}
\end{tikzpicture}
    \caption{Two-family pattern}
    \label{fig:tri-duo2}
  \end{subfigure}
  \hspace{1em}
  \begin{subfigure}[t]{0.45\linewidth}
    \centering
    \begin{tikzpicture}[scale=0.75]
  \clip (-4.0000,-4.0000) rectangle (4.0000,4.0000);
  \begin{scope}[lightgray]
    \draw (-8.7500,-7.5777) -- (8.7500,-7.5777);
    \draw (-8.7500,-6.0622) -- (8.7500,-6.0622);
    \draw (-8.7500,-4.5466) -- (8.7500,-4.5466);
    \draw (-8.7500,-3.0311) -- (8.7500,-3.0311);
    \draw (-8.7500,-1.5155) -- (8.7500,-1.5155);
    \draw (-8.7500,0) -- (8.7500,0);
    \draw (-8.7500,1.5155) -- (8.7500,1.5155);
    \draw (-8.7500,3.0311) -- (8.7500,3.0311);
    \draw (-8.7500,4.5466) -- (8.7500,4.5466);
    \draw (-8.7500,6.0622) -- (8.7500,6.0622);
    \draw (-8.7500,7.5777) -- (8.7500,7.5777);
    \draw (-16.625,-7.5777) -- (-7.8750,7.5777);
    \draw (-14.875,-7.5777) -- (-6.1250,7.5777);
    \draw (-13.125,-7.5777) -- (-4.3750,7.5777);
    \draw (-11.375,-7.5777) -- (-2.6250,7.5777);
    \draw (-9.6250,-7.5777) -- (-0.87500,7.5777);
    \draw (-7.8750,-7.5777) -- (0.87500,7.5777);
    \draw (-6.1250,-7.5777) -- (2.6250,7.5777);
    \draw (-4.3750,-7.5777) -- (4.3750,7.5777);
    \draw (-2.6250,-7.5777) -- (6.1250,7.5777);
    \draw (-0.87500,-7.5777) -- (7.8750,7.5777);
    \draw (0.87500,-7.5777) -- (9.6250,7.5777);
    \draw (2.6250,-7.5777) -- (11.375,7.5777);
    \draw (4.3750,-7.5777) -- (13.125,7.5777);
    \draw (6.1250,-7.5777) -- (14.875,7.5777);
    \draw (7.8750,-7.5777) -- (16.625,7.5777);
    \draw (-7.8750,-7.5777) -- (-16.625,7.5777);
    \draw (-6.1250,-7.5777) -- (-14.875,7.5777);
    \draw (-4.3750,-7.5777) -- (-13.125,7.5777);
    \draw (-2.6250,-7.5777) -- (-11.375,7.5777);
    \draw (-0.87500,-7.5777) -- (-9.6250,7.5777);
    \draw (0.87500,-7.5777) -- (-7.8750,7.5777);
    \draw (2.6250,-7.5777) -- (-6.1250,7.5777);
    \draw (4.3750,-7.5777) -- (-4.3750,7.5777);
    \draw (6.1250,-7.5777) -- (-2.6250,7.5777);
    \draw (7.8750,-7.5777) -- (-0.87500,7.5777);
    \draw (9.6250,-7.5777) -- (0.87500,7.5777);
    \draw (11.375,-7.5777) -- (2.6250,7.5777);
    \draw (13.125,-7.5777) -- (4.3750,7.5777);
    \draw (14.875,-7.5777) -- (6.1250,7.5777);
    \draw (16.625,-7.5777) -- (7.8750,7.5777);
  \end{scope}
  \begin{scope}[red, thick, fill opacity=0.1]
    \filldraw (-4.3750,-1.5155) circle[radius=1];
    \filldraw (-0.87500,-4.5466) circle[radius=1];
    \filldraw (-3.5000,3.0311) circle[radius=1];
    \filldraw (0,0) circle[radius=1];
    \filldraw (3.5000,-3.0311) circle[radius=1];
    \filldraw (0.87500,4.5466) circle[radius=1];
    \filldraw (4.3750,1.5155) circle[radius=1];
  \end{scope}
  \begin{scope}[blue, thick, fill opacity=0.1]
    \filldraw (-2.6250,-3.0311) circle[radius=1];
    \filldraw (-1.7500,1.5155) circle[radius=1];
    \filldraw (1.7500,-1.5155) circle[radius=1];
    \filldraw (2.6250,3.0311) circle[radius=1];
  \end{scope}
  \begin{scope}[green!70!black, thick, fill opacity=0.1]
    \filldraw (-2.1875,-0.75777) circle[radius=1];
    \filldraw (1.3125,-3.7889) circle[radius=1];
    \filldraw (-1.3125,3.7889) circle[radius=1];
    \filldraw (2.1875,0.75777) circle[radius=1];
  \end{scope}
  \begin{scope}[orange!80!olive, thick, fill opacity=0.1]
    \filldraw (-4.8125,-3.7889) circle[radius=1];
    \filldraw (-3.9375,0.75777) circle[radius=1];
    \filldraw (-0.43750,-2.2733) circle[radius=1];
    \filldraw (0.43750,2.2733) circle[radius=1];
    \filldraw (3.9375,-0.75777) circle[radius=1];
    \filldraw (4.8125,3.7889) circle[radius=1];
  \end{scope}
  \begin{scope}[red, fill opacity=0.1]
  \end{scope}
  \begin{scope}[red!60!black, thick]
    \fill (-4.3750,-1.5155) circle[radius=0.080000];
    \fill (-0.87500,-4.5466) circle[radius=0.080000];
    \fill (-3.5000,3.0311) circle[radius=0.080000];
    \fill (0,0) circle[radius=0.080000];
    \fill (3.5000,-3.0311) circle[radius=0.080000];
    \fill (0.87500,4.5466) circle[radius=0.080000];
    \fill (4.3750,1.5155) circle[radius=0.080000];
  \end{scope}
  \begin{scope}[blue, fill opacity=0.1]
  \end{scope}
  \begin{scope}[blue!60!black, thick]
    \draw (-3.5000,-3.0311) -- (-1.7500,-3.0311);
    \fill (-3.5000,-3.0311) circle[radius=0.080000];
    \fill (-1.7500,-3.0311) circle[radius=0.080000];
    \draw (-0.87500,1.5155) -- (-2.6250,1.5155);
    \fill (-0.87500,1.5155) circle[radius=0.080000];
    \fill (-2.6250,1.5155) circle[radius=0.080000];
    \draw (0.87500,-1.5155) -- (2.6250,-1.5155);
    \fill (0.87500,-1.5155) circle[radius=0.080000];
    \fill (2.6250,-1.5155) circle[radius=0.080000];
    \draw (1.7500,3.0311) -- (3.5000,3.0311);
    \fill (1.7500,3.0311) circle[radius=0.080000];
    \fill (3.5000,3.0311) circle[radius=0.080000];
  \end{scope}
  \begin{scope}[green!70!black, fill opacity=0.1]
  \end{scope}
  \begin{scope}[green!70!black!60!black, thick]
    \draw (-2.6250,-1.5155) -- (-1.7500,0);
    \fill (-2.6250,-1.5155) circle[radius=0.080000];
    \fill (-1.7500,0) circle[radius=0.080000];
    \draw (1.7500,-3.0311) -- (0.87500,-4.5466);
    \fill (1.7500,-3.0311) circle[radius=0.080000];
    \fill (0.87500,-4.5466) circle[radius=0.080000];
    \draw (-1.7500,3.0311) -- (-0.87500,4.5466);
    \fill (-1.7500,3.0311) circle[radius=0.080000];
    \fill (-0.87500,4.5466) circle[radius=0.080000];
    \draw (1.7500,0) -- (2.6250,1.5155);
    \fill (1.7500,0) circle[radius=0.080000];
    \fill (2.6250,1.5155) circle[radius=0.080000];
  \end{scope}
  \begin{scope}[orange!80!olive, fill opacity=0.1]
  \end{scope}
  \begin{scope}[orange!80!olive!60!black, thick]
    \draw (-5.2500,-3.0311) -- (-4.3750,-4.5466);
    \fill (-5.2500,-3.0311) circle[radius=0.080000];
    \fill (-4.3750,-4.5466) circle[radius=0.080000];
    \draw (-3.5000,0) -- (-4.3750,1.5155);
    \fill (-3.5000,0) circle[radius=0.080000];
    \fill (-4.3750,1.5155) circle[radius=0.080000];
    \draw (-0.87500,-1.5155) -- (0,-3.0311);
    \fill (-0.87500,-1.5155) circle[radius=0.080000];
    \fill (0,-3.0311) circle[radius=0.080000];
    \draw (0.87500,1.5155) -- (0,3.0311);
    \fill (0.87500,1.5155) circle[radius=0.080000];
    \fill (0,3.0311) circle[radius=0.080000];
    \draw (4.3750,-1.5155) -- (3.5000,0);
    \fill (4.3750,-1.5155) circle[radius=0.080000];
    \fill (3.5000,0) circle[radius=0.080000];
    \draw (5.2500,3.0311) -- (4.3750,4.5466);
    \fill (5.2500,3.0311) circle[radius=0.080000];
    \fill (4.3750,4.5466) circle[radius=0.080000];
  \end{scope}
\end{tikzpicture}
    \caption{Four-family pattern}
    \label{fig:tri-mono1trio2}
  \end{subfigure}
  \caption{Two motif patterns for the two-point regime, both valid for $\frac{4}{\sqrt7} \le d \le 2$.}
  \label{fig:tri-pair-patterns}
\end{figure}

For the main proof, we use the two simpler patterns shown in \cref{fig:tri-pair-patterns}. Both are valid for
\[
  \frac{4}{\sqrt7} \le d \le 2 .
\]
The first pattern consists of two families of two-point motifs, with the two families using different orientations of lattice edges. The second pattern uses four families: three two-point families, one in each edge orientation of the triangular mesh, together with one single-point family. At the lower endpoint, the realizing disks in the four-family pattern form a hexagonal packing.

Combining these two-point patterns with the primitive patterns gives a covering for every
\[
  d \in \left[\frac{2}{\sqrt7},\infty\right).
\]

\subsection{Single-family motif patterns}
\label{sec:tri-single-family}

We next describe additional single-family motif patterns. The motifs responsible for $1$ and $7$ lattice points in \cref{fig:tri-uniform1,fig:tri-uniform7} share an important feature: the center of each realizing disk lies on a lattice point. The same principle extends to motifs containing larger neighborhoods of a lattice point.

\begin{figure}[!t]
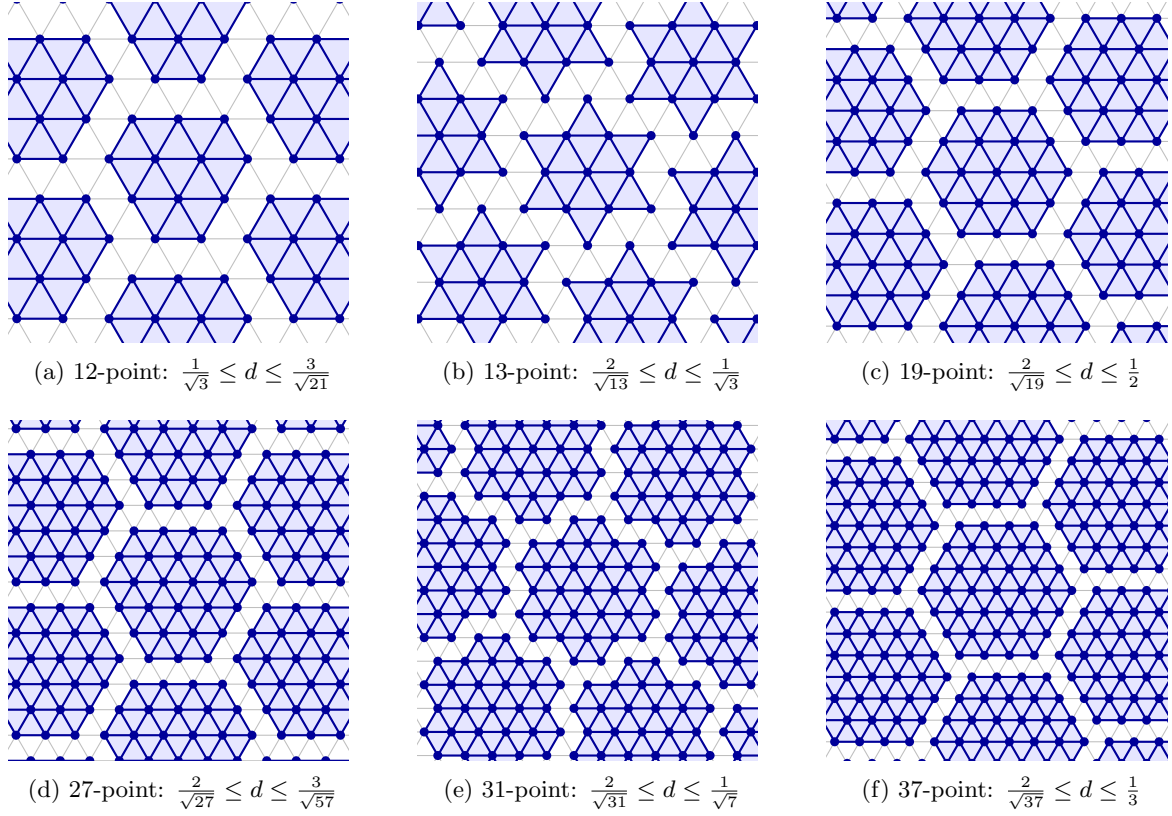

  \centering
  \begin{subfigure}[t]{0.32\linewidth}
    \centering
    \input{img/tri/uniform12.tikz}
    \caption{12-point: $\frac{1}{\sqrt3} \le d \le \frac{3}{\sqrt{21}}$}
    \label{fig:tri-uniform12}
  \end{subfigure}
  \hfill
  \begin{subfigure}[t]{0.32\linewidth}
    \centering
    \input{img/tri/uniform13.tikz}
    \caption{13-point: $\frac{2}{\sqrt{13}} \le d \le \frac{1}{\sqrt3}$}
    \label{fig:tri-uniform13}
  \end{subfigure}
  \hfill
  \begin{subfigure}[t]{0.32\linewidth}
    \centering
    \input{img/tri/uniform19.tikz}
    \caption{19-point: $\frac{2}{\sqrt{19}} \le d \le \frac{1}{2}$}
    \label{fig:tri-uniform19}
  \end{subfigure}
  \par\vspace{1em}
  \begin{subfigure}[t]{0.32\linewidth}
    \centering
    \input{img/tri/uniform27.tikz}
    \caption{27-point: $\frac{2}{\sqrt{27}} \le d \le \frac{3}{\sqrt{57}}$}
    \label{fig:tri-uniform27}
  \end{subfigure}
  \hfill
  \begin{subfigure}[t]{0.32\linewidth}
    \centering
    \input{img/tri/uniform31.tikz}
    \caption{31-point: $\frac{2}{\sqrt{31}} \le d \le \frac{1}{\sqrt7}$}
    \label{fig:tri-uniform31}
  \end{subfigure}
  \hfill
  \begin{subfigure}[t]{0.32\linewidth}
    \centering
    \input{img/tri/uniform37.tikz}
    \caption{37-point: $\frac{2}{\sqrt{37}} \le d \le \frac{1}{3}$}
    \label{fig:tri-uniform37}
  \end{subfigure}
  \caption{Single-family motif patterns whose realizing disks are centered at lattice vertices or mesh faces.}
  \label{fig:tri-vertex-face}
\end{figure}

Let
\[
  Q(a,b) = a^2-ab+b^2
\]
be the squared norm of the Eisenstein integer $a+b\omega$. For several values of $r$, we take the motif consisting of all lattice points satisfying $Q(a,b)\le r$, with its realizing disk centered at the origin. The cases $r=3,4,7,9$ give motifs with $13$, $19$, $31$, and $37$ lattice points, respectively. Suitable translation lattices then produce the single-family patterns shown in \cref{fig:tri-uniform13,fig:tri-uniform19,fig:tri-uniform31,fig:tri-uniform37}.

An analogous construction begins with the triangle pattern in \cref{fig:tri-uniform3}, whose realizing disk is centered at a face of the triangular mesh. Enlarging the assigned neighborhood about such a face yields motifs with $12$ and $27$ lattice points; the resulting patterns are shown in \cref{fig:tri-uniform12,fig:tri-uniform27}. As in the vertex-based cases, the lower endpoint for each pattern is determined by the minimum separation between realizing centers, while the upper endpoint is determined by the farthest assigned lattice point.

\begin{figure}[h]
  \centering
  \begin{subfigure}[t]{0.45\linewidth}
    \centering
    \begin{tikzpicture}[scale=2.5]
  \clip (-1.4760,-1.2000) rectangle (0.92400,1.2000);
  \begin{scope}[lightgray]
    \draw (-3.4500,-2.9878) -- (3.4500,-2.9878);
    \draw (-3.4500,-2.3902) -- (3.4500,-2.3902);
    \draw (-3.4500,-1.7927) -- (3.4500,-1.7927);
    \draw (-3.4500,-1.1951) -- (3.4500,-1.1951);
    \draw (-3.4500,-0.59756) -- (3.4500,-0.59756);
    \draw (-3.4500,0) -- (3.4500,0);
    \draw (-3.4500,0.59756) -- (3.4500,0.59756);
    \draw (-3.4500,1.1951) -- (3.4500,1.1951);
    \draw (-3.4500,1.7927) -- (3.4500,1.7927);
    \draw (-3.4500,2.3902) -- (3.4500,2.3902);
    \draw (-3.4500,2.9878) -- (3.4500,2.9878);
    \draw (-6.5550,-2.9878) -- (-3.1050,2.9878);
    \draw (-5.8650,-2.9878) -- (-2.4150,2.9878);
    \draw (-5.1750,-2.9878) -- (-1.7250,2.9878);
    \draw (-4.4850,-2.9878) -- (-1.0350,2.9878);
    \draw (-3.7950,-2.9878) -- (-0.34500,2.9878);
    \draw (-3.1050,-2.9878) -- (0.34500,2.9878);
    \draw (-2.4150,-2.9878) -- (1.0350,2.9878);
    \draw (-1.7250,-2.9878) -- (1.7250,2.9878);
    \draw (-1.0350,-2.9878) -- (2.4150,2.9878);
    \draw (-0.34500,-2.9878) -- (3.1050,2.9878);
    \draw (0.34500,-2.9878) -- (3.7950,2.9878);
    \draw (1.0350,-2.9878) -- (4.4850,2.9878);
    \draw (1.7250,-2.9878) -- (5.1750,2.9878);
    \draw (2.4150,-2.9878) -- (5.8650,2.9878);
    \draw (3.1050,-2.9878) -- (6.5550,2.9878);
    \draw (-3.1050,-2.9878) -- (-6.5550,2.9878);
    \draw (-2.4150,-2.9878) -- (-5.8650,2.9878);
    \draw (-1.7250,-2.9878) -- (-5.1750,2.9878);
    \draw (-1.0350,-2.9878) -- (-4.4850,2.9878);
    \draw (-0.34500,-2.9878) -- (-3.7950,2.9878);
    \draw (0.34500,-2.9878) -- (-3.1050,2.9878);
    \draw (1.0350,-2.9878) -- (-2.4150,2.9878);
    \draw (1.7250,-2.9878) -- (-1.7250,2.9878);
    \draw (2.4150,-2.9878) -- (-1.0350,2.9878);
    \draw (3.1050,-2.9878) -- (-0.34500,2.9878);
    \draw (3.7950,-2.9878) -- (0.34500,2.9878);
    \draw (4.4850,-2.9878) -- (1.0350,2.9878);
    \draw (5.1750,-2.9878) -- (1.7250,2.9878);
    \draw (5.8650,-2.9878) -- (2.4150,2.9878);
    \draw (6.5550,-2.9878) -- (3.1050,2.9878);
  \end{scope}
  \begin{scope}[blue, thick, fill opacity=0.05]
    \filldraw (-1.3110,-1.7927) circle[radius=1];
    \filldraw (0.75900,-1.7927) circle[radius=1];
    \filldraw (-2.3460,0) circle[radius=1];
    \filldraw (-0.27600,0) circle[radius=1];
    \filldraw (1.7940,0) circle[radius=1];
    \filldraw (-1.3110,1.7927) circle[radius=1];
    \filldraw (0.75900,1.7927) circle[radius=1];
  \end{scope}
  \begin{scope}[blue, fill opacity=0.1]
    \fill (-2.0700,-1.1951) -- (-1.7250,-1.7927) -- (-2.0700,-2.3902) -- (-0.69000,-2.3902) -- (-0.34500,-1.7927) -- (-0.69000,-1.1951) -- cycle;
    \fill (0,-1.1951) -- (0.34500,-1.7927) -- (0,-2.3902) -- (1.3800,-2.3902) -- (1.7250,-1.7927) -- (1.3800,-1.1951) -- cycle;
    \fill (-1.7250,-0.59756) -- (-1.3800,0) -- (-1.7250,0.59756) -- (-3.1050,0.59756) -- (-2.7600,0) -- (-3.1050,-0.59756) -- cycle;
    \fill (-1.0350,-0.59756) -- (0.34500,-0.59756) -- (0.69000,0) -- (0.34500,0.59756) -- (-1.0350,0.59756) -- (-0.69000,0) -- cycle;
    \fill (1.0350,-0.59756) -- (2.4150,-0.59756) -- (2.7600,0) -- (2.4150,0.59756) -- (1.0350,0.59756) -- (1.3800,0) -- cycle;
    \fill (-0.69000,1.1951) -- (-0.34500,1.7927) -- (-0.69000,2.3902) -- (-2.0700,2.3902) -- (-1.7250,1.7927) -- (-2.0700,1.1951) -- cycle;
    \fill (0,1.1951) -- (1.3800,1.1951) -- (1.7250,1.7927) -- (1.3800,2.3902) -- (0,2.3902) -- (0.34500,1.7927) -- cycle;
  \end{scope}
  \begin{scope}[blue!60!black, thick]
    \draw (-2.0700,-1.1951) -- (-1.7250,-1.7927) -- (-2.0700,-2.3902) -- (-0.69000,-2.3902) -- (-0.34500,-1.7927) -- (-0.69000,-1.1951) -- cycle;
    \fill (-1.3800,-1.1951) circle[radius=0.024000];
    \fill (-2.0700,-1.1951) circle[radius=0.024000];
    \fill (-0.69000,-1.1951) circle[radius=0.024000];
    \fill (-1.7250,-1.7927) circle[radius=0.024000];
    \fill (-1.0350,-1.7927) circle[radius=0.024000];
    \fill (-0.34500,-1.7927) circle[radius=0.024000];
    \fill (-2.0700,-2.3902) circle[radius=0.024000];
    \fill (-1.3800,-2.3902) circle[radius=0.024000];
    \fill (-0.69000,-2.3902) circle[radius=0.024000];
    \draw (0,-1.1951) -- (0.34500,-1.7927) -- (0,-2.3902) -- (1.3800,-2.3902) -- (1.7250,-1.7927) -- (1.3800,-1.1951) -- cycle;
    \fill (0,-1.1951) circle[radius=0.024000];
    \fill (0.69000,-1.1951) circle[radius=0.024000];
    \fill (1.3800,-1.1951) circle[radius=0.024000];
    \fill (0.34500,-1.7927) circle[radius=0.024000];
    \fill (1.0350,-1.7927) circle[radius=0.024000];
    \fill (1.7250,-1.7927) circle[radius=0.024000];
    \fill (0,-2.3902) circle[radius=0.024000];
    \fill (0.69000,-2.3902) circle[radius=0.024000];
    \fill (1.3800,-2.3902) circle[radius=0.024000];
    \draw (-1.7250,-0.59756) -- (-1.3800,0) -- (-1.7250,0.59756) -- (-3.1050,0.59756) -- (-2.7600,0) -- (-3.1050,-0.59756) -- cycle;
    \fill (-1.7250,-0.59756) circle[radius=0.024000];
    \fill (-2.4150,-0.59756) circle[radius=0.024000];
    \fill (-3.1050,-0.59756) circle[radius=0.024000];
    \fill (-1.3800,0) circle[radius=0.024000];
    \fill (-2.0700,0) circle[radius=0.024000];
    \fill (-2.7600,0) circle[radius=0.024000];
    \fill (-1.7250,0.59756) circle[radius=0.024000];
    \fill (-2.4150,0.59756) circle[radius=0.024000];
    \fill (-3.1050,0.59756) circle[radius=0.024000];
    \draw (-1.0350,-0.59756) -- (0.34500,-0.59756) -- (0.69000,0) -- (0.34500,0.59756) -- (-1.0350,0.59756) -- (-0.69000,0) -- cycle;
    \fill (-1.0350,-0.59756) circle[radius=0.024000];
    \fill (-0.34500,-0.59756) circle[radius=0.024000];
    \fill (0.34500,-0.59756) circle[radius=0.024000];
    \fill (-0.69000,0) circle[radius=0.024000];
    \fill (0,0) circle[radius=0.024000];
    \fill (0.69000,0) circle[radius=0.024000];
    \fill (-0.34500,0.59756) circle[radius=0.024000];
    \fill (-1.0350,0.59756) circle[radius=0.024000];
    \fill (0.34500,0.59756) circle[radius=0.024000];
    \draw (1.0350,-0.59756) -- (2.4150,-0.59756) -- (2.7600,0) -- (2.4150,0.59756) -- (1.0350,0.59756) -- (1.3800,0) -- cycle;
    \fill (1.0350,-0.59756) circle[radius=0.024000];
    \fill (1.7250,-0.59756) circle[radius=0.024000];
    \fill (2.4150,-0.59756) circle[radius=0.024000];
    \fill (1.3800,0) circle[radius=0.024000];
    \fill (2.0700,0) circle[radius=0.024000];
    \fill (2.7600,0) circle[radius=0.024000];
    \fill (1.0350,0.59756) circle[radius=0.024000];
    \fill (1.7250,0.59756) circle[radius=0.024000];
    \fill (2.4150,0.59756) circle[radius=0.024000];
    \draw (-0.69000,1.1951) -- (-0.34500,1.7927) -- (-0.69000,2.3902) -- (-2.0700,2.3902) -- (-1.7250,1.7927) -- (-2.0700,1.1951) -- cycle;
    \fill (-0.69000,1.1951) circle[radius=0.024000];
    \fill (-1.3800,1.1951) circle[radius=0.024000];
    \fill (-2.0700,1.1951) circle[radius=0.024000];
    \fill (-0.34500,1.7927) circle[radius=0.024000];
    \fill (-1.0350,1.7927) circle[radius=0.024000];
    \fill (-1.7250,1.7927) circle[radius=0.024000];
    \fill (-0.69000,2.3902) circle[radius=0.024000];
    \fill (-1.3800,2.3902) circle[radius=0.024000];
    \fill (-2.0700,2.3902) circle[radius=0.024000];
    \draw (0,1.1951) -- (1.3800,1.1951) -- (1.7250,1.7927) -- (1.3800,2.3902) -- (0,2.3902) -- (0.34500,1.7927) -- cycle;
    \fill (0,1.1951) circle[radius=0.024000];
    \fill (0.69000,1.1951) circle[radius=0.024000];
    \fill (1.3800,1.1951) circle[radius=0.024000];
    \fill (0.34500,1.7927) circle[radius=0.024000];
    \fill (1.0350,1.7927) circle[radius=0.024000];
    \fill (1.7250,1.7927) circle[radius=0.024000];
    \fill (0.69000,2.3902) circle[radius=0.024000];
    \fill (0,2.3902) circle[radius=0.024000];
    \fill (1.3800,2.3902) circle[radius=0.024000];
  \end{scope}
  \begin{scope}[blue!60!black, thick]
    \draw (-1.3110,-1.7927) circle[radius=0.036000];
    \draw (0.75900,-1.7927) circle[radius=0.036000];
    \draw (-2.3460,0) circle[radius=0.036000];
    \draw (-0.27600,0) circle[radius=0.036000];
    \draw (1.7940,0) circle[radius=0.036000];
    \draw (-1.3110,1.7927) circle[radius=0.036000];
    \draw (0.75900,1.7927) circle[radius=0.036000];
  \end{scope}
  \begin{scope}[-stealth, black, thick, font=\small]
    \draw (0,0) -- (-0.276,0) node[midway, anchor=north] {$\frac25d$};
  \end{scope}
\end{tikzpicture}
    \caption{9-point: $\frac23 \le d \le \frac57$}
    \label{fig:tri-uniform9-zoom}
  \end{subfigure}
  \hspace{1em}
  \begin{subfigure}[t]{0.45\linewidth}
    \centering
    \input{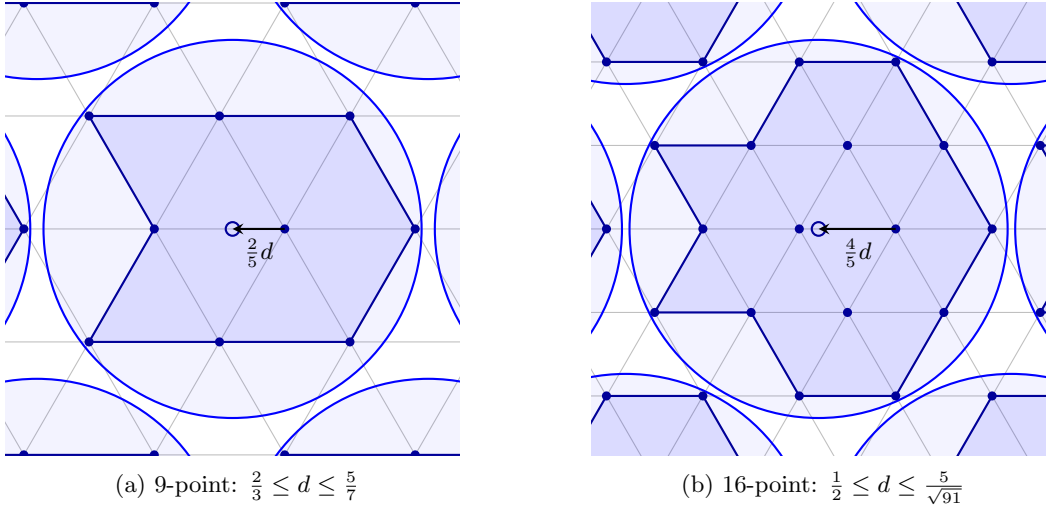}
    \caption{16-point: $\frac12 \le d \le \frac5{\sqrt{91}}$}
    \label{fig:tri-uniform16-zoom}
  \end{subfigure}
  \caption{Single-family motif patterns with off-lattice realizing centers.}
  \label{fig:tri-off-lattice}
\end{figure}

Single-family patterns can also be obtained when the realizing centers lie neither at lattice points nor at face centers. Consider the 9-point motif
\[
  M_9
  =
  \big\{ a+b\omega : a \in \{-1,0,1\},\ b \in \{0,1\} \big\}
  \cup
  \big\{ a-\omega : a \in \{-2,-1,0\} \big\}.
\]
A realizing disk for this motif is centered at the Chebyshev center, which is shifted from the origin by $\frac25d$ in the negative real direction. The farthest points of the motif are at distance $\frac75d$ from this center. Hence the containment condition gives $d\le 5/7$. The period translations in \cref{fig:tri-uniform9-zoom} have shortest center separation $3d$, so non-overlap gives $d\ge 2/3$. Therefore, this motif pattern is valid for
\[
  \frac23 \le d \le \frac57.
\]

The same approach gives a 16-point motif whose realizing center is chosen to minimize the maximum distance to its assigned points. For the translation lattice shown in \cref{fig:tri-uniform16-zoom}, the non-overlap and containment constraints give
\[
  \frac12 \le d \le \frac5{\sqrt{91}}.
\]

\subsection{Multi-family motif patterns}
\label{sec:tri-multifamily}

The four-family pattern in \cref{fig:tri-mono1trio2} shows that different motif families can be combined while preserving an overall hexagonal-packing structure for the realizing disks. This observation leads to further multi-family constructions.

\begin{figure}[!b]
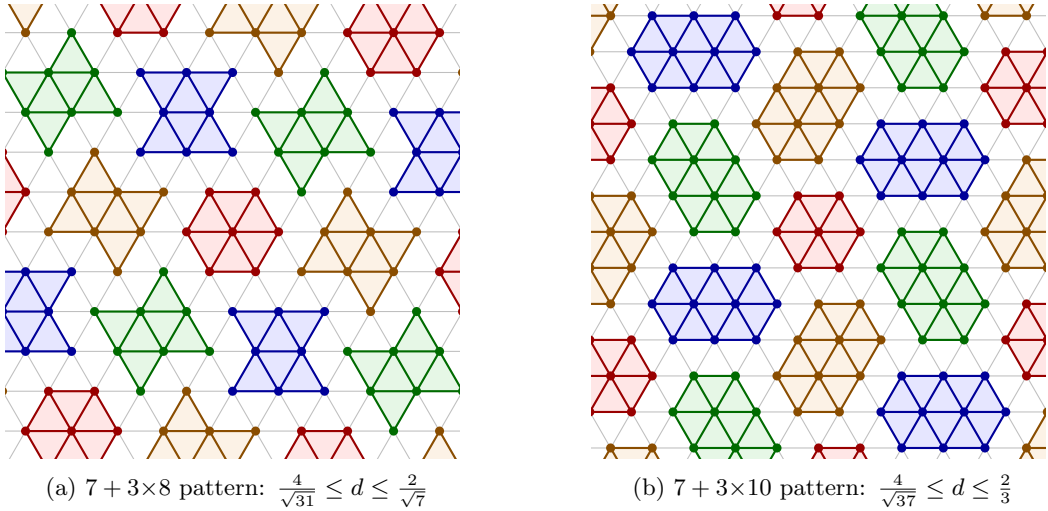

  \centering
  \begin{subfigure}[t]{0.45\linewidth}
    \centering
    \input{img/tri/mono7trio8.tikz}
    \caption{$7+3{\times}8$ pattern: $\frac{4}{\sqrt{31}} \le d \le \frac{2}{\sqrt7}$}
    \label{fig:tri-mono7trio8}
  \end{subfigure}
  \hspace{1em}
  \begin{subfigure}[t]{0.45\linewidth}
    \centering
    \input{img/tri/mono7trio10.tikz}
    \caption{$7+3{\times}10$ pattern: $\frac{4}{\sqrt{37}} \le d \le \frac23$}
    \label{fig:tri-mono7trio10}
  \end{subfigure}
  \caption{Multi-family motif patterns for the triangular lattice.}
  \label{fig:tri-multifamily}
\end{figure}

The first such construction is a four-family motif pattern. One family is responsible for $7$ lattice points and is realized by disks centered at lattice points. The other three families are each responsible for $8$ lattice points and are realized by disks centered at midpoints of edges of the triangular mesh. See \cref{fig:tri-mono7trio8}. The four families together are responsible for
\[
  7+8+8+8 = 31
\]
lattice points per tile. At the lower endpoint, the realizing disks form a hexagonal packing, giving the non-overlap bound
\[
  d \ge \frac{4}{\sqrt{31}}.
\]
The upper endpoint is determined by the most restrictive containment condition among the four families, which gives
\[
  d \le \frac{2}{\sqrt7}.
\]
Thus, this construction is valid for
\[
  \frac{4}{\sqrt{31}} \le d \le \frac{2}{\sqrt7}.
\]

The second construction has the same four-family structure, but enlarges each edge-based motif from $8$ to $10$ lattice points; see \cref{fig:tri-mono7trio10}. Thus, one family is responsible for $7$ lattice points per motif and the other three families are each responsible for $10$, giving
\[
  7+10+10+10 = 37
\]
lattice points per tile. The density bound is attained when the realizing disks form a hexagonal packing, which gives
\[
  d \ge \frac4{\sqrt{37}}.
\]
The edge-based motifs determine the upper endpoint through their containment constraint, giving $d\le 2/3$. Hence this construction is valid for
\[
  \frac4{\sqrt{37}} \le d \le \frac23.
\]

\begin{remark}[Density bound]\label{rmk:tri-density}
  The lower endpoints of many of the constructions follow from a common density calculation. Suppose that a periodic pattern uses $m$ disks to cover $K$ lattice points in each fundamental tile. Thus, each disk is responsible for $K/m$ points on average. If the realizing centers have the density of a triangular lattice of spacing $\alpha d$, then comparison with the point density $2/(\sqrt3d^2)$ of $\mathcal{T}_d$ gives
  \[
    \frac2{\sqrt3(\alpha d)^2}
    =
    \frac1{K/m} \cdot \frac2{\sqrt3d^2},
  \]
  and hence $\alpha^2=K/m$. Non-overlap requires $\alpha d\ge2$, so necessarily
  \[
    d \ge \frac2{\sqrt{K/m}} = 2\sqrt{\frac{m}{K}}.
  \]
  Equality occurs when the realizing disks form a hexagonal packing. In particular, for a single-family pattern whose disks are each responsible for $k$ points, the bound reduces to $d\ge2/\sqrt{k}$; for a four-family pattern covering $K$ points per fundamental tile, it becomes $d\ge4/\sqrt K$.
\end{remark}

\section{Honeycomb Point Set}
\label{sec:hex}

The honeycomb point set is a subset of the triangular lattice. Therefore, any covering of $\mathcal{T}_d$ by pairwise non-overlapping unit disks immediately yields a covering of $\mathcal{H}_d$ for the same spacing $d$. In particular, every spacing interval established in \cref{thm:tri-spacing} for the triangular lattice also applies to the honeycomb point set.

Nevertheless, the honeycomb point set admits several additional motif patterns that are more natural than those obtained by restriction from the triangular lattice. Some of these simplify the inherited constructions, while others enlarge the range of valid spacings. Combining the inherited intervals with these additional patterns gives the following result.

\begin{theorem}\label{thm:hex-spacing}
  The honeycomb point set $\mathcal{H}_d$ can be covered by pairwise non-overlapping unit disks whenever
  \[
    d \in
    \left[ \frac2{\sqrt{37}} , \frac1{\sqrt7} \right]
    \cup
    \left[ \frac2{\sqrt{27}} , \frac3{\sqrt{57}} \right]
    \cup
    \left[ \frac2{\sqrt{21}} , \frac5{\sqrt{91}} \right]
    \cup
    \left[ \frac2{\sqrt{13}} , \frac3{\sqrt{21}} \right]
    \cup
    \left[ \frac4{\sqrt{37}} , \infty \right).
  \]
\end{theorem}

Thus, the gaps
\[
  \frac13 < d < \frac2{\sqrt{31}}
  \qquad\text{and}\qquad
  \frac57 < d < \frac4{\sqrt{31}}
\]
left by the triangular-lattice constructions are filled in the honeycomb setting.

For reference, \cref{tab:hex-patterns} summarizes the honeycomb motif patterns used in this section. The pair, hexagon, 14-point, and 24-point patterns are native to the honeycomb point set. The alternating-singleton and partial-rhombus patterns are included to illustrate how triangular-lattice constructions restrict to the honeycomb point set; the remaining inherited patterns are omitted.

\begin{table}[b]
  \centering
  \small
  \begin{tabular}{lcccccc}
    \toprule
    \multirow{2}{*}{Pattern}
      & \multirow{2}{*}{Centers}
      & \multirow{2}{*}{Responsibilities}
      & \multicolumn{2}{c}{$d_{\min}$} & \multicolumn{2}{c}{$d_{\max}$} \\
    \cmidrule(lr){4-5}\cmidrule(lr){6-7}
      & & & Exact & Approx. & Exact & Approx. \\
    \midrule
    Alt.\ singleton
      & vertices
      & $1+1$
      & $2$
      & $2.000$
      & $\infty$
      & - \\

    Pair
      & edge
      & $2$
      & $2/\sqrt3$
      & $1.155$
      & $2$
      & $2.000$ \\

    Part.\ rhombus
      & Chebyshev
      & $2+3+3$
      & $1$
      & $1.000$
      & $2/\sqrt3$
      & $1.155$ \\

    Hexagon
      & face
      & $6$
      & $2/3$
      & $0.667$
      & $1$
      & $1.000$ \\

    Single-family
      & edge
      & $14$
      & $2/\sqrt{21}$
      & $0.436$
      & $2/\sqrt{19}$
      & $0.459$ \\

    Single-family
      & face
      & $24$
      & $1/3$
      & $0.333$
      & $1/\sqrt7$
      & $0.378$ \\
    \bottomrule
  \end{tabular}
  \caption{Periodic motif patterns used for the honeycomb point set.}
  \label{tab:hex-patterns}
\end{table}

\begin{figure}[!b]
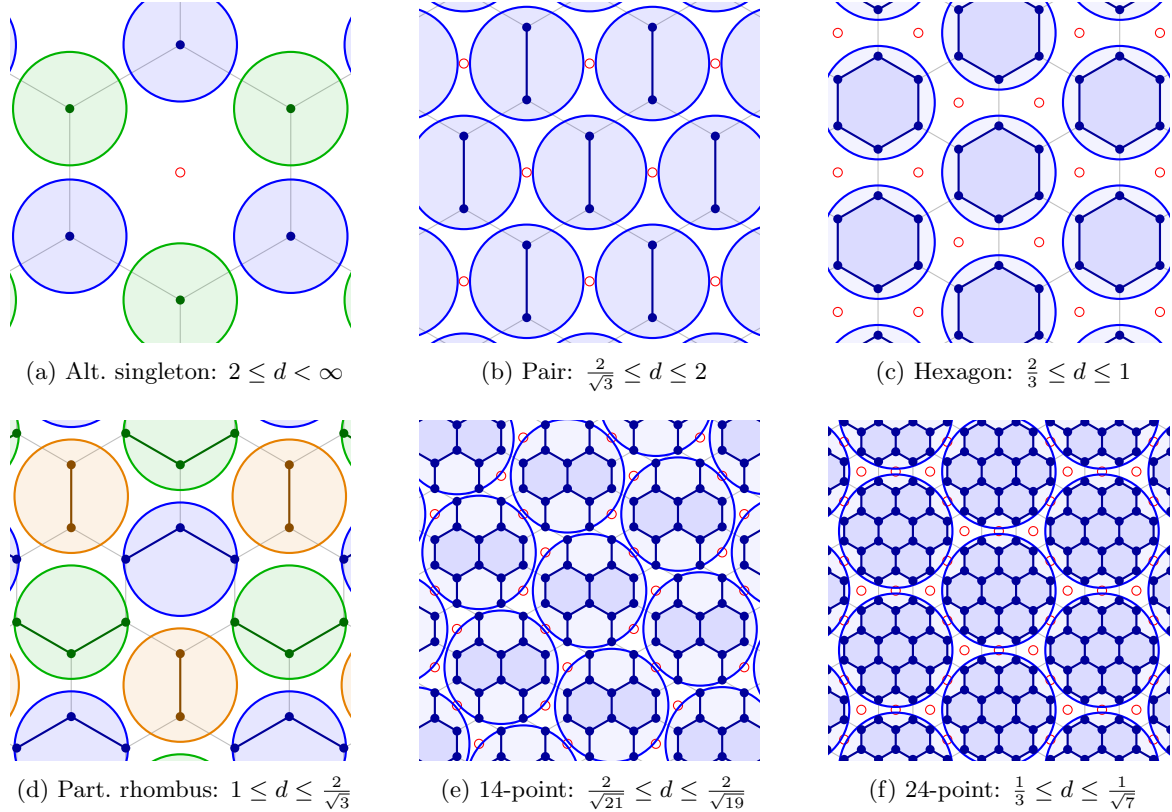

  \centering
  \begin{subfigure}[t]{0.32\linewidth}
    \centering
    \begin{tikzpicture}[scale=0.75]
  \clip (-3.0000,-3.0000) rectangle (3.0000,3.0000);
  \begin{scope}[lightgray]
    \draw (0,-4.5000) -- (0,-2.2500);
    \draw (1.9486,-1.1250) -- (0,-2.2500);
    \draw (1.9486,1.1250) -- (0,2.2500);
    \draw (-3.8971,-2.2500) -- (-1.9486,-1.1250);
    \draw (-3.8971,2.2500) -- (-1.9486,1.1250);
    \draw (1.9486,-1.1250) -- (3.8971,-2.2500);
    \draw (3.8971,2.2500) -- (1.9486,1.1250);
    \draw (-1.9486,1.1250) -- (0,2.2500);
    \draw (-1.9486,1.1250) -- (-1.9486,-1.1250);
    \draw (0,-2.2500) -- (-1.9486,-1.1250);
    \draw (0,4.5000) -- (0,2.2500);
    \draw (1.9486,-1.1250) -- (1.9486,1.1250);
  \end{scope}
  \begin{scope}[red]
    \draw (0,0) circle[radius=0.080000];
  \end{scope}
  \begin{scope}[green!70!black, thick, fill opacity=0.1]
    \filldraw (-3.8971,-2.2500) circle[radius=1];
    \filldraw (0,-2.2500) circle[radius=1];
    \filldraw (3.8971,-2.2500) circle[radius=1];
    \filldraw (-1.9486,1.1250) circle[radius=1];
    \filldraw (1.9486,1.1250) circle[radius=1];
  \end{scope}
  \begin{scope}[blue, thick, fill opacity=0.1]
    \filldraw (-1.9486,-1.1250) circle[radius=1];
    \filldraw (1.9486,-1.1250) circle[radius=1];
    \filldraw (-3.8971,2.2500) circle[radius=1];
    \filldraw (0,2.2500) circle[radius=1];
    \filldraw (3.8971,2.2500) circle[radius=1];
  \end{scope}
  \begin{scope}[green!70!black, fill opacity=0.1]
  \end{scope}
  \begin{scope}[blue, fill opacity=0.1]
  \end{scope}
  \begin{scope}[green!70!black!60!black, thick]
    \fill (-3.8971,-2.2500) circle[radius=0.080000];
    \fill (0,-2.2500) circle[radius=0.080000];
    \fill (3.8971,-2.2500) circle[radius=0.080000];
    \fill (-1.9486,1.1250) circle[radius=0.080000];
    \fill (1.9486,1.1250) circle[radius=0.080000];
  \end{scope}
  \begin{scope}[blue!60!black, thick]
    \fill (-1.9486,-1.1250) circle[radius=0.080000];
    \fill (1.9486,-1.1250) circle[radius=0.080000];
    \fill (-3.8971,2.2500) circle[radius=0.080000];
    \fill (0,2.2500) circle[radius=0.080000];
    \fill (3.8971,2.2500) circle[radius=0.080000];
  \end{scope}
\end{tikzpicture}
    \caption{Alt.\ singleton: $2 \le d < \infty$}
    \label{fig:hex-duo1}
  \end{subfigure}
  \hfill
  \begin{subfigure}[t]{0.32\linewidth}
    \centering
    \begin{tikzpicture}[scale=0.75]
  \clip (-1.8915,-3.0000) rectangle (4.1085,3.0000);
  \begin{scope}[lightgray]
    \draw (1.1085,-0.64000) -- (2.2170,-1.2800);
    \draw (2.2170,1.2800) -- (3.3255,0.64000);
    \draw (2.2170,1.2800) -- (1.1085,0.64000);
    \draw (-2.2170,1.2800) -- (-1.1085,0.64000);
    \draw (3.3255,3.2000) -- (2.2170,2.5600);
    \draw (3.3255,-0.64000) -- (4.4341,-1.2800);
    \draw (-1.1085,-3.2000) -- (-2.2170,-2.5600);
    \draw (0,-1.2800) -- (-1.1085,-0.64000);
    \draw (-2.2170,-1.2800) -- (-1.1085,-0.64000);
    \draw (3.3255,3.2000) -- (4.4341,2.5600);
    \draw (3.3255,-3.2000) -- (4.4341,-2.5600);
    \draw (-1.1085,0.64000) -- (0,1.2800);
    \draw (2.2170,1.2800) -- (2.2170,2.5600);
    \draw (0,2.5600) -- (0,1.2800);
    \draw (-1.1085,0.64000) -- (-1.1085,-0.64000);
    \draw (1.1085,3.2000) -- (2.2170,2.5600);
    \draw (2.2170,-2.5600) -- (3.3255,-3.2000);
    \draw (0,-2.5600) -- (1.1085,-3.2000);
    \draw (0,2.5600) -- (1.1085,3.2000);
    \draw (1.1085,0.64000) -- (0,1.2800);
    \draw (0,-2.5600) -- (-1.1085,-3.2000);
    \draw (0,-2.5600) -- (0,-1.2800);
    \draw (1.1085,-0.64000) -- (0,-1.2800);
    \draw (3.3255,-0.64000) -- (3.3255,0.64000);
    \draw (2.2170,-2.5600) -- (1.1085,-3.2000);
    \draw (2.2170,-2.5600) -- (2.2170,-1.2800);
    \draw (3.3255,-0.64000) -- (2.2170,-1.2800);
    \draw (1.1085,-0.64000) -- (1.1085,0.64000);
    \draw (4.4341,1.2800) -- (3.3255,0.64000);
    \draw (0,2.5600) -- (-1.1085,3.2000);
    \draw (-2.2170,2.5600) -- (-1.1085,3.2000);
  \end{scope}
  \begin{scope}[red]
    \draw (-1.1085,-1.9200) circle[radius=0.080000];
    \draw (1.1085,-1.9200) circle[radius=0.080000];
    \draw (3.3255,-1.9200) circle[radius=0.080000];
    \draw (0,0) circle[radius=0.080000];
    \draw (2.2170,0) circle[radius=0.080000];
    \draw (-1.1085,1.9200) circle[radius=0.080000];
    \draw (1.1085,1.9200) circle[radius=0.080000];
    \draw (3.3255,1.9200) circle[radius=0.080000];
  \end{scope}
  \begin{scope}[blue, thick, fill opacity=0.1]
    \filldraw (-1.1085,-3.8400) circle[radius=1];
    \filldraw (1.1085,-3.8400) circle[radius=1];
    \filldraw (3.3255,-3.8400) circle[radius=1];
    \filldraw (-2.2170,-1.9200) circle[radius=1];
    \filldraw (0,-1.9200) circle[radius=1];
    \filldraw (2.2170,-1.9200) circle[radius=1];
    \filldraw (4.4341,-1.9200) circle[radius=1];
    \filldraw (-1.1085,0) circle[radius=1];
    \filldraw (1.1085,0) circle[radius=1];
    \filldraw (3.3255,0) circle[radius=1];
    \filldraw (-2.2170,1.9200) circle[radius=1];
    \filldraw (0,1.9200) circle[radius=1];
    \filldraw (2.2170,1.9200) circle[radius=1];
    \filldraw (4.4341,1.9200) circle[radius=1];
    \filldraw (-1.1085,3.8400) circle[radius=1];
    \filldraw (1.1085,3.8400) circle[radius=1];
    \filldraw (3.3255,3.8400) circle[radius=1];
  \end{scope}
  \begin{scope}[blue, fill opacity=0.1]
  \end{scope}
  \begin{scope}[blue!60!black, thick]
    \fill (-1.1085,-3.2000) circle[radius=0.080000];
    \fill (1.1085,-3.2000) circle[radius=0.080000];
    \fill (3.3255,-3.2000) circle[radius=0.080000];
    \draw (-2.2170,-2.5600) -- (-2.2170,-1.2800);
    \fill (-2.2170,-1.2800) circle[radius=0.080000];
    \fill (-2.2170,-2.5600) circle[radius=0.080000];
    \draw (0,-2.5600) -- (0,-1.2800);
    \fill (0,-1.2800) circle[radius=0.080000];
    \fill (0,-2.5600) circle[radius=0.080000];
    \draw (2.2170,-2.5600) -- (2.2170,-1.2800);
    \fill (2.2170,-1.2800) circle[radius=0.080000];
    \fill (2.2170,-2.5600) circle[radius=0.080000];
    \draw (4.4341,-2.5600) -- (4.4341,-1.2800);
    \fill (4.4341,-1.2800) circle[radius=0.080000];
    \fill (4.4341,-2.5600) circle[radius=0.080000];
    \draw (-1.1085,0.64000) -- (-1.1085,-0.64000);
    \fill (-1.1085,-0.64000) circle[radius=0.080000];
    \fill (-1.1085,0.64000) circle[radius=0.080000];
    \draw (1.1085,0.64000) -- (1.1085,-0.64000);
    \fill (1.1085,-0.64000) circle[radius=0.080000];
    \fill (1.1085,0.64000) circle[radius=0.080000];
    \draw (3.3255,0.64000) -- (3.3255,-0.64000);
    \fill (3.3255,-0.64000) circle[radius=0.080000];
    \fill (3.3255,0.64000) circle[radius=0.080000];
    \draw (-2.2170,2.5600) -- (-2.2170,1.2800);
    \fill (-2.2170,1.2800) circle[radius=0.080000];
    \fill (-2.2170,2.5600) circle[radius=0.080000];
    \draw (0,2.5600) -- (0,1.2800);
    \fill (0,1.2800) circle[radius=0.080000];
    \fill (0,2.5600) circle[radius=0.080000];
    \draw (2.2170,2.5600) -- (2.2170,1.2800);
    \fill (2.2170,1.2800) circle[radius=0.080000];
    \fill (2.2170,2.5600) circle[radius=0.080000];
    \draw (4.4341,2.5600) -- (4.4341,1.2800);
    \fill (4.4341,1.2800) circle[radius=0.080000];
    \fill (4.4341,2.5600) circle[radius=0.080000];
    \fill (-1.1085,3.2000) circle[radius=0.080000];
    \fill (1.1085,3.2000) circle[radius=0.080000];
    \fill (3.3255,3.2000) circle[radius=0.080000];
  \end{scope}
\end{tikzpicture}
    \caption{Pair: $\frac{2}{\sqrt3} \le d \le 2$}
    \label{fig:hex-uniform2}
  \end{subfigure}
  \hfill
  \begin{subfigure}[t]{0.32\linewidth}
    \centering
    \input{img/hex/uniform6.tikz}
    \caption{Hexagon: $\frac23 \le d \le 1$}
    \label{fig:hex-uniform6}
  \end{subfigure}
  \par\vspace{1em}
  \begin{subfigure}[t]{0.32\linewidth}
    \centering
    \begin{tikzpicture}[scale=0.75]
  \clip (-3.0000,-3.0000) rectangle (3.0000,3.0000);
  \begin{scope}[lightgray]
    \draw (2.8839,0.55500) -- (3.8452,1.1100);
    \draw (0.96129,-2.7750) -- (1.9226,-2.2200);
    \draw (1.9226,-1.1100) -- (1.9226,-2.2200);
    \draw (1.9226,-1.1100) -- (2.8839,-0.55500);
    \draw (-0.96129,0.55500) -- (0,1.1100);
    \draw (2.8839,-2.7750) -- (2.8839,-3.8850);
    \draw (-3.8452,2.2200) -- (-2.8839,2.7750);
    \draw (1.9226,1.1100) -- (1.9226,2.2200);
    \draw (-2.8839,-0.55500) -- (-2.8839,0.55500);
    \draw (2.8839,2.7750) -- (2.8839,3.8850);
    \draw (-2.8839,0.55500) -- (-3.8452,1.1100);
    \draw (-0.96129,0.55500) -- (-1.9226,1.1100);
    \draw (-1.9226,1.1100) -- (-2.8839,0.55500);
    \draw (-2.8839,-2.7750) -- (-1.9226,-2.2200);
    \draw (-1.9226,-1.1100) -- (-1.9226,-2.2200);
    \draw (-1.9226,-1.1100) -- (-0.96129,-0.55500);
    \draw (0.96129,-2.7750) -- (0,-2.2200);
    \draw (1.9226,-1.1100) -- (0.96129,-0.55500);
    \draw (0,-2.2200) -- (-0.96129,-2.7750);
    \draw (0,-2.2200) -- (0,-1.1100);
    \draw (0.96129,-0.55500) -- (0,-1.1100);
    \draw (0.96129,3.8850) -- (0.96129,2.7750);
    \draw (-2.8839,-2.7750) -- (-2.8839,-3.8850);
    \draw (2.8839,-2.7750) -- (1.9226,-2.2200);
    \draw (3.8452,-1.1100) -- (2.8839,-0.55500);
    \draw (2.8839,0.55500) -- (2.8839,-0.55500);
    \draw (0.96129,2.7750) -- (0,2.2200);
    \draw (0.96129,-2.7750) -- (0.96129,-3.8850);
    \draw (2.8839,2.7750) -- (1.9226,2.2200);
    \draw (0.96129,-0.55500) -- (0.96129,0.55500);
    \draw (-2.8839,-2.7750) -- (-3.8452,-2.2200);
    \draw (-1.9226,-1.1100) -- (-2.8839,-0.55500);
    \draw (-2.8839,-0.55500) -- (-3.8452,-1.1100);
    \draw (-0.96129,-2.7750) -- (-0.96129,-3.8850);
    \draw (-1.9226,-2.2200) -- (-0.96129,-2.7750);
    \draw (-0.96129,-0.55500) -- (0,-1.1100);
    \draw (1.9226,1.1100) -- (2.8839,0.55500);
    \draw (1.9226,1.1100) -- (0.96129,0.55500);
    \draw (-0.96129,3.8850) -- (-0.96129,2.7750);
    \draw (-2.8839,3.8850) -- (-2.8839,2.7750);
    \draw (2.8839,-2.7750) -- (3.8452,-2.2200);
    \draw (3.8452,2.2200) -- (2.8839,2.7750);
    \draw (-2.8839,2.7750) -- (-1.9226,2.2200);
    \draw (-0.96129,2.7750) -- (0,2.2200);
    \draw (-1.9226,2.2200) -- (-0.96129,2.7750);
    \draw (-1.9226,1.1100) -- (-1.9226,2.2200);
    \draw (0,1.1100) -- (0.96129,0.55500);
    \draw (0,2.2200) -- (0,1.1100);
    \draw (-0.96129,0.55500) -- (-0.96129,-0.55500);
    \draw (1.9226,2.2200) -- (0.96129,2.7750);
  \end{scope}
  \begin{scope}[orange!80!olive, thick, fill opacity=0.1]
    \filldraw (-3.8452,-1.6650) circle[radius=1];
    \filldraw (-1.9226,1.6650) circle[radius=1];
    \filldraw (0,-1.6650) circle[radius=1];
    \filldraw (1.9226,1.6650) circle[radius=1];
    \filldraw (3.8452,-1.6650) circle[radius=1];
  \end{scope}
  \begin{scope}[blue, thick, fill opacity=0.1]
    \filldraw (-3.8452,0.55500) circle[radius=1];
    \filldraw (-1.9226,-2.7750) circle[radius=1];
    \filldraw (-1.9226,3.8850) circle[radius=1];
    \filldraw (0,0.55500) circle[radius=1];
    \filldraw (1.9226,-2.7750) circle[radius=1];
    \filldraw (1.9226,3.8850) circle[radius=1];
    \filldraw (3.8452,0.55500) circle[radius=1];
  \end{scope}
  \begin{scope}[green!70!black, thick, fill opacity=0.1]
    \filldraw (-3.8452,-3.8850) circle[radius=1];
    \filldraw (-3.8452,2.7750) circle[radius=1];
    \filldraw (-1.9226,-0.55500) circle[radius=1];
    \filldraw (0,-3.8850) circle[radius=1];
    \filldraw (0,2.7750) circle[radius=1];
    \filldraw (1.9226,-0.55500) circle[radius=1];
    \filldraw (3.8452,-3.8850) circle[radius=1];
    \filldraw (3.8452,2.7750) circle[radius=1];
  \end{scope}
  \begin{scope}[orange!80!olive, fill opacity=0.1]
  \end{scope}
  \begin{scope}[blue, fill opacity=0.1]
  \end{scope}
  \begin{scope}[green!70!black, fill opacity=0.1]
  \end{scope}
  \begin{scope}[orange!80!olive!60!black, thick]
    \draw (-3.8452,-2.2200) -- (-3.8452,-1.1100);
    \fill (-3.8452,-1.1100) circle[radius=0.080000];
    \fill (-3.8452,-2.2200) circle[radius=0.080000];
    \draw (-1.9226,2.2200) -- (-1.9226,1.1100);
    \fill (-1.9226,1.1100) circle[radius=0.080000];
    \fill (-1.9226,2.2200) circle[radius=0.080000];
    \draw (0,-2.2200) -- (0,-1.1100);
    \fill (0,-1.1100) circle[radius=0.080000];
    \fill (0,-2.2200) circle[radius=0.080000];
    \draw (1.9226,2.2200) -- (1.9226,1.1100);
    \fill (1.9226,1.1100) circle[radius=0.080000];
    \fill (1.9226,2.2200) circle[radius=0.080000];
    \draw (3.8452,-2.2200) -- (3.8452,-1.1100);
    \fill (3.8452,-1.1100) circle[radius=0.080000];
    \fill (3.8452,-2.2200) circle[radius=0.080000];
  \end{scope}
  \begin{scope}[blue!60!black, thick]
    \draw (-2.8839,0.55500) -- (-3.8452,1.1100);
    \fill (-3.8452,1.1100) circle[radius=0.080000];
    \fill (-2.8839,0.55500) circle[radius=0.080000];
    \draw (-1.9226,-2.2200) -- (-2.8839,-2.7750);
    \draw (-0.96129,-2.7750) -- (-1.9226,-2.2200);
    \fill (-2.8839,-2.7750) circle[radius=0.080000];
    \fill (-1.9226,-2.2200) circle[radius=0.080000];
    \fill (-0.96129,-2.7750) circle[radius=0.080000];
    \fill (-0.96129,3.8850) circle[radius=0.080000];
    \fill (-2.8839,3.8850) circle[radius=0.080000];
    \draw (0,1.1100) -- (-0.96129,0.55500);
    \draw (0.96129,0.55500) -- (0,1.1100);
    \fill (-0.96129,0.55500) circle[radius=0.080000];
    \fill (0,1.1100) circle[radius=0.080000];
    \fill (0.96129,0.55500) circle[radius=0.080000];
    \draw (1.9226,-2.2200) -- (0.96129,-2.7750);
    \draw (2.8839,-2.7750) -- (1.9226,-2.2200);
    \fill (0.96129,-2.7750) circle[radius=0.080000];
    \fill (1.9226,-2.2200) circle[radius=0.080000];
    \fill (2.8839,-2.7750) circle[radius=0.080000];
    \fill (0.96129,3.8850) circle[radius=0.080000];
    \fill (2.8839,3.8850) circle[radius=0.080000];
    \draw (3.8452,1.1100) -- (2.8839,0.55500);
    \fill (2.8839,0.55500) circle[radius=0.080000];
    \fill (3.8452,1.1100) circle[radius=0.080000];
  \end{scope}
  \begin{scope}[green!70!black!60!black, thick]
    \fill (-2.8839,-3.8850) circle[radius=0.080000];
    \draw (-3.8452,2.2200) -- (-2.8839,2.7750);
    \fill (-2.8839,2.7750) circle[radius=0.080000];
    \fill (-3.8452,2.2200) circle[radius=0.080000];
    \draw (-1.9226,-1.1100) -- (-2.8839,-0.55500);
    \draw (-0.96129,-0.55500) -- (-1.9226,-1.1100);
    \fill (-2.8839,-0.55500) circle[radius=0.080000];
    \fill (-1.9226,-1.1100) circle[radius=0.080000];
    \fill (-0.96129,-0.55500) circle[radius=0.080000];
    \fill (-0.96129,-3.8850) circle[radius=0.080000];
    \fill (0.96129,-3.8850) circle[radius=0.080000];
    \draw (0,2.2200) -- (-0.96129,2.7750);
    \draw (0.96129,2.7750) -- (0,2.2200);
    \fill (-0.96129,2.7750) circle[radius=0.080000];
    \fill (0,2.2200) circle[radius=0.080000];
    \fill (0.96129,2.7750) circle[radius=0.080000];
    \draw (1.9226,-1.1100) -- (0.96129,-0.55500);
    \draw (2.8839,-0.55500) -- (1.9226,-1.1100);
    \fill (0.96129,-0.55500) circle[radius=0.080000];
    \fill (1.9226,-1.1100) circle[radius=0.080000];
    \fill (2.8839,-0.55500) circle[radius=0.080000];
    \fill (2.8839,-3.8850) circle[radius=0.080000];
    \draw (3.8452,2.2200) -- (2.8839,2.7750);
    \fill (2.8839,2.7750) circle[radius=0.080000];
    \fill (3.8452,2.2200) circle[radius=0.080000];
  \end{scope}
\end{tikzpicture}
    \caption{Part.\ rhombus: $1 \le d \le \frac2{\sqrt3}$}
    \label{fig:hex-mono2duo3}
  \end{subfigure}
  \hfill
  \begin{subfigure}[t]{0.32\linewidth}
    \centering
    \input{img/hex/uniform14.tikz}
    \caption{14-point: $\frac2{\sqrt{21}} \le d \le \frac2{\sqrt{19}}$}
    \label{fig:hex-uniform14}
  \end{subfigure}
  \hfill
  \begin{subfigure}[t]{0.32\linewidth}
    \centering
    \input{img/hex/uniform24.tikz}
    \caption{24-point: $\frac13 \le d \le \frac1{\sqrt7}$}
    \label{fig:hex-uniform24}
  \end{subfigure}
  \caption{Motif patterns with realizations for the honeycomb point set. Red points indicate triangular-lattice points deleted in forming the honeycomb point set.}
  \label{fig:hex-patterns}
\end{figure}

We first consider two patterns inherited from triangular-lattice constructions. Recall from \cref{sec:prelim} that $\mathcal H$ is obtained by partitioning $\mathcal T$ into three congruence classes and deleting one class. Restricting the singleton pattern in \cref{fig:tri-uniform1}, valid for $d\ge 2$, therefore leaves two families of singleton motifs, one for each retained class. This gives the alternating-singleton pattern in \cref{fig:hex-duo1}.

Such a reduction in the number of families is not always possible: at smaller spacings, every family of disks in the restricted construction may remain responsible for points of $\mathcal H_d$. For example, restricting the triangular-lattice rhombus pattern in \cref{fig:tri-uniform4} yields the three-family partial-rhombus pattern in \cref{fig:hex-mono2duo3}, valid for
\[
  1 \le d \le \frac2{\sqrt3}.
\]
The three motif families are responsible for $2$, $3$, and $3$ honeycomb points, respectively, and none can be discarded. Although the realizing centers have simple descriptions relative to the triangular lattice, their coordinates are less natural relative to the honeycomb geometry.

We next give four native single-family patterns. The first is the pair pattern in \cref{fig:hex-uniform2}, in which each motif consists of two adjacent points of $\mathcal H_d$. It has a valid realization for
\[
  \frac2{\sqrt3} \le d \le 2.
\]
Thus, this single honeycomb-native construction replaces the triangular-lattice triangle and four-family patterns shown in \cref{fig:tri-uniform3,fig:tri-mono1trio2} over the relevant spacing range.

The second is the hexagon pattern shown in \cref{fig:hex-uniform6}, in which each disk is responsible for six points around a face of the honeycomb mesh. This gives a valid covering for
\[
  \frac23 \le d \le 1.
\]
Together with the inherited constructions on either side, this pattern closes the gap
\[
  \frac57 < d < \frac4{\sqrt{31}}.
\]

The third is the 14-point pattern in \cref{fig:hex-uniform14}. Each realizing disk is centered at the midpoint of an edge of the honeycomb mesh. The pattern is valid for
\[
  \frac2{\sqrt{21}} \le d \le \frac2{\sqrt{19}}.
\]
It therefore narrows the gap between the inherited 27-point and 19-point intervals from
\[
  \frac3{\sqrt{57}} < d < \frac2{\sqrt{19}}
  \qquad\text{to}\qquad
  \frac3{\sqrt{57}} < d < \frac2{\sqrt{21}}.
\]

Finally, the 24-point pattern in \cref{fig:hex-uniform24} assigns to each disk a neighborhood of a honeycomb face. The period translations have length $6d$, so pairwise non-overlap requires $6d\ge2$. The farthest assigned points lie at distance $\sqrt7d$ from the realizing center, giving the containment constraint $\sqrt7d\le1$. Hence this pattern is valid for
\[
  \frac13 \le d \le \frac1{\sqrt7}.
\]
Its interval joins the vertex-based 37-point interval at $d=1/3$ and overlaps the inherited 31-point interval, thereby closing the gap between them.

Taking the union of these additional intervals with those inherited from \cref{thm:tri-spacing} proves \cref{thm:hex-spacing}.

\section{Square Lattice}
\label{sec:square}

The square lattice was previously studied by Alm et~al.~\cite{alm2013motif}. They gave several motif constructions, in addition to the trivial singleton construction for $d\ge 2$, and stated the following coverability result.

\begin{claim}[Alm et~al.]\label{thm:square-alm}
  The square lattice $\mathcal{S}_d$ can be covered by pairwise non-overlapping unit disks whenever
  \[
    d \in
    \left[ \frac2{\sqrt{13}} , \frac1{\sqrt2} \right]
    \cup
    \left[ \frac4{\sqrt{26}} , \infty \right).
  \]
\end{claim}

However, one of their motif constructions used to establish the claim, intended for the interval
\[
  \frac2{\sqrt{10}} \le d \le \frac23,
\]
is not valid: in the proposed realization, some disks overlap. The remaining constructions of Alm et~al.\ are valid and provide the starting point for this section.

We give two additional constructions. The first is a modification of the invalid construction of Alm et~al.; it recovers part of the missing interval. The second is a new ``zipper'' pattern, which gives a coverability interval at smaller spacings. Together with the valid constructions of Alm et~al., these imply the following theorem.

\begin{theorem}\label{thm:square-spacing}
  The square lattice $\mathcal{S}_d$ can be covered by pairwise non-overlapping unit disks whenever
  \[
    d \in
    \left[ \frac12 , \frac6{\sqrt{130}} \right]
    \cup
    \left[ \frac2{\sqrt{13}} , \frac2{\sqrt{10}} \right]
    \cup
    \left[ \frac4{\sqrt{37}} , \frac1{\sqrt2} \right]
    \cup
    \left[ \frac4{\sqrt{26}} , \infty \right).
  \]
\end{theorem}

For reference, \cref{tab:square-patterns} summarizes the motif patterns used in this section. The singleton pattern is included as the trivial baseline. Patterns marked with an asterisk are nontrivial constructions of Alm et~al.

\begin{table}[b]
  \centering
  \small
  \begin{tabular}{lcccccc}
    \toprule
    \multirow{2}{*}{Pattern}
      & \multirow{2}{*}{Centers}
      & \multirow{2}{*}{Responsibilities}
      & \multicolumn{2}{c}{$d_{\min}$} & \multicolumn{2}{c}{$d_{\max}$} \\
    \cmidrule(lr){4-5}\cmidrule(lr){6-7}
      & & & Exact & Approx. & Exact & Approx. \\
    \midrule
    Singleton
      & vertex
      & $1$
      & $2$
      & $2.000$
      & $\infty$
      & - \\

    Pair*
      & edge
      & $2$
      & $\sqrt2$
      & $1.414$
      & $2$
      & $2.000$ \\

    Square*
      & face
      & $4$
      & $1$
      & $1.000$
      & $\sqrt2$
      & $1.414$ \\

    Cross*
      & vertex
      & $5$
      & $2/\sqrt5$
      & $0.894$
      & $1$
      & $1.000$ \\

    Alt.\ rectangle*
      & edges
      & $6+6$
      & $4/\sqrt{26}$
      & $0.784$
      & $2/\sqrt5$
      & $0.894$ \\

    Two-family
      & edge and vertex
      & $8+9$
      & $4/\sqrt{37}$
      & $0.658$
      & $2/3$
      & $0.667$ \\

    Big square*
      & vertex
      & $9$
      & $2/3$
      & $0.667$
      & $1/\sqrt2$
      & $0.707$ \\

    Big cross*
      & face
      & $12$
      & $2/\sqrt{13}$
      & $0.555$
      & $2/\sqrt{10}$
      & $0.632$ \\

    Zipper
      & variable
      & $14+14$
      & $1/2$
      & $0.500$
      & $6/\sqrt{130}$
      & $0.526$ \\
    \bottomrule
    \multicolumn{7}{l}{\small Patterns marked with * are nontrivial patterns appearing in Alm et~al.}
  \end{tabular}
  \caption{Periodic motif patterns for the square lattice.}
  \label{tab:square-patterns}
\end{table}

\subsection{Motif patterns of Alm et~al.}
\label{sec:square-alm-patterns}

\begin{figure}[!t]
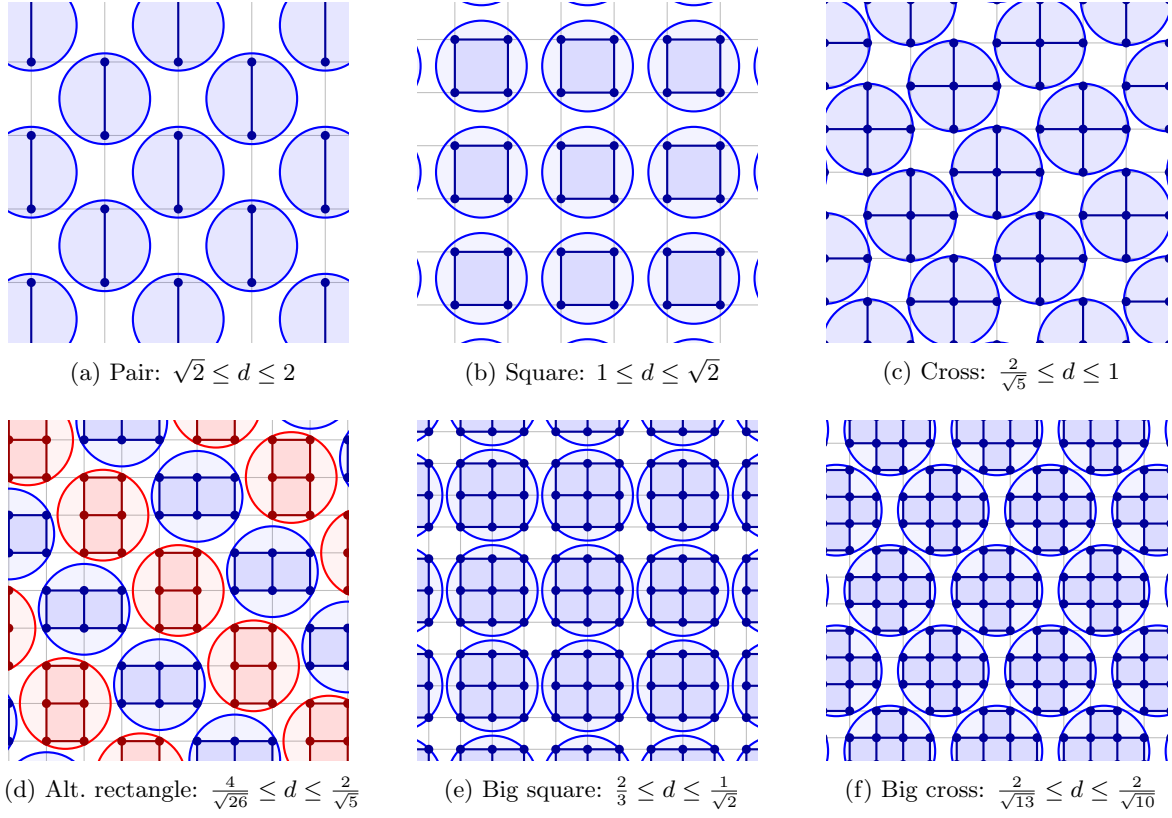

  \centering
  \begin{subfigure}[t]{0.32\linewidth}
    \centering
    \begin{tikzpicture}[scale=0.6]
  \clip (-3.7500,-2.9400) rectangle (3.7500,4.5600);
  \begin{scope}[lightgray]
    \draw (-8.1000,-8.1000) -- (8.1000,-8.1000);
    \draw (-8.1000,-6.4800) -- (8.1000,-6.4800);
    \draw (-8.1000,-4.8600) -- (8.1000,-4.8600);
    \draw (-8.1000,-3.2400) -- (8.1000,-3.2400);
    \draw (-8.1000,-1.6200) -- (8.1000,-1.6200);
    \draw (-8.1000,0) -- (8.1000,0);
    \draw (-8.1000,1.6200) -- (8.1000,1.6200);
    \draw (-8.1000,3.2400) -- (8.1000,3.2400);
    \draw (-8.1000,4.8600) -- (8.1000,4.8600);
    \draw (-8.1000,6.4800) -- (8.1000,6.4800);
    \draw (-8.1000,8.1000) -- (8.1000,8.1000);
    \draw (-8.1000,-8.1000) -- (-8.1000,8.1000);
    \draw (-6.4800,-8.1000) -- (-6.4800,8.1000);
    \draw (-4.8600,-8.1000) -- (-4.8600,8.1000);
    \draw (-3.2400,-8.1000) -- (-3.2400,8.1000);
    \draw (-1.6200,-8.1000) -- (-1.6200,8.1000);
    \draw (0,-8.1000) -- (0,8.1000);
    \draw (1.6200,-8.1000) -- (1.6200,8.1000);
    \draw (3.2400,-8.1000) -- (3.2400,8.1000);
    \draw (4.8600,-8.1000) -- (4.8600,8.1000);
    \draw (6.4800,-8.1000) -- (6.4800,8.1000);
    \draw (8.1000,-8.1000) -- (8.1000,8.1000);
  \end{scope}
  \begin{scope}[blue, thick, fill opacity=0.1]
    \filldraw (-3.2400,-2.4300) circle[radius=1];
    \filldraw (-3.2400,0.81000) circle[radius=1];
    \filldraw (-3.2400,4.0500) circle[radius=1];
    \filldraw (-1.6200,-0.81000) circle[radius=1];
    \filldraw (-1.6200,2.4300) circle[radius=1];
    \filldraw (0,-2.4300) circle[radius=1];
    \filldraw (0,0.81000) circle[radius=1];
    \filldraw (0,4.0500) circle[radius=1];
    \filldraw (1.6200,-0.81000) circle[radius=1];
    \filldraw (1.6200,2.4300) circle[radius=1];
    \filldraw (3.2400,-2.4300) circle[radius=1];
    \filldraw (3.2400,0.81000) circle[radius=1];
    \filldraw (3.2400,4.0500) circle[radius=1];
  \end{scope}
  \begin{scope}[blue, fill opacity=0.1]
  \end{scope}
  \begin{scope}[blue!60!black, thick]
    \draw (-3.2400,-1.6200) -- (-3.2400,-3.2400);
    \fill (-3.2400,-1.6200) circle[radius=0.10000];
    \fill (-3.2400,-3.2400) circle[radius=0.10000];
    \draw (-3.2400,0) -- (-3.2400,1.6200);
    \fill (-3.2400,0) circle[radius=0.10000];
    \fill (-3.2400,1.6200) circle[radius=0.10000];
    \draw (-3.2400,3.2400) -- (-3.2400,4.8600);
    \fill (-3.2400,3.2400) circle[radius=0.10000];
    \fill (-3.2400,4.8600) circle[radius=0.10000];
    \draw (-1.6200,-1.6200) -- (-1.6200,0);
    \fill (-1.6200,-1.6200) circle[radius=0.10000];
    \fill (-1.6200,0) circle[radius=0.10000];
    \draw (-1.6200,1.6200) -- (-1.6200,3.2400);
    \fill (-1.6200,1.6200) circle[radius=0.10000];
    \fill (-1.6200,3.2400) circle[radius=0.10000];
    \draw (0,-1.6200) -- (0,-3.2400);
    \fill (0,-1.6200) circle[radius=0.10000];
    \fill (0,-3.2400) circle[radius=0.10000];
    \draw (0,0) -- (0,1.6200);
    \fill (0,0) circle[radius=0.10000];
    \fill (0,1.6200) circle[radius=0.10000];
    \draw (0,3.2400) -- (0,4.8600);
    \fill (0,3.2400) circle[radius=0.10000];
    \fill (0,4.8600) circle[radius=0.10000];
    \draw (1.6200,-1.6200) -- (1.6200,0);
    \fill (1.6200,-1.6200) circle[radius=0.10000];
    \fill (1.6200,0) circle[radius=0.10000];
    \draw (1.6200,1.6200) -- (1.6200,3.2400);
    \fill (1.6200,1.6200) circle[radius=0.10000];
    \fill (1.6200,3.2400) circle[radius=0.10000];
    \draw (3.2400,-1.6200) -- (3.2400,-3.2400);
    \fill (3.2400,-1.6200) circle[radius=0.10000];
    \fill (3.2400,-3.2400) circle[radius=0.10000];
    \draw (3.2400,0) -- (3.2400,1.6200);
    \fill (3.2400,0) circle[radius=0.10000];
    \fill (3.2400,1.6200) circle[radius=0.10000];
    \draw (3.2400,3.2400) -- (3.2400,4.8600);
    \fill (3.2400,3.2400) circle[radius=0.10000];
    \fill (3.2400,4.8600) circle[radius=0.10000];
  \end{scope}
\end{tikzpicture}
    \caption{Pair: $\sqrt2 \le d \le 2$}
    \label{fig:square-uniform2}
  \end{subfigure}
  \hfill
  \begin{subfigure}[t]{0.32\linewidth}
    \centering
    \input{img/square/uniform4.tikz}
    \caption{Square: $1 \le d \le \sqrt2$}
    \label{fig:square-uniform4}
  \end{subfigure}
  \hfill
  \begin{subfigure}[t]{0.32\linewidth}
    \centering
    \input{img/square/uniform5.tikz}
    \caption{Cross: $\frac{2}{\sqrt5} \le d \le 1$}
    \label{fig:square-uniform5}
  \end{subfigure}
  \par\vspace{1em}
  \begin{subfigure}[t]{0.32\linewidth}
    \centering
    \input{img/square/duo6.tikz}
    \caption{Alt.\ rectangle: $\frac{4}{\sqrt{26}} \le d \le \frac{2}{\sqrt5}$}
    \label{fig:square-duo6}
  \end{subfigure}
  \hfill
  \begin{subfigure}[t]{0.32\linewidth}
    \centering
    \input{img/square/uniform9.tikz}
    \caption{Big square: $\frac23 \le d \le \frac{1}{\sqrt2}$}
    \label{fig:square-uniform9}
  \end{subfigure}
  \hfill
  \begin{subfigure}[t]{0.32\linewidth}
    \centering
    \input{img/square/uniform12.tikz}
    \caption{Big cross: $\frac{2}{\sqrt{13}} \le d \le \frac{2}{\sqrt{10}}$}
    \label{fig:square-uniform12}
  \end{subfigure}
  \caption{Six valid nontrivial motif patterns of Alm et~al.\ for the square lattice.}
  \label{fig:square-alm}
\end{figure}

Alm et~al.\ construct their motif patterns using a single motif together with translations and rotations that preserve the square lattice. In our translation-only framework, all but one of their valid constructions remain single-family motif patterns. The alternating-rectangle construction becomes a two-family motif pattern. The six valid nontrivial patterns are shown in \cref{fig:square-alm}.

\begin{figure}[!b]
  \centering
  \begin{subfigure}[t]{0.45\linewidth}
    \centering
    \input{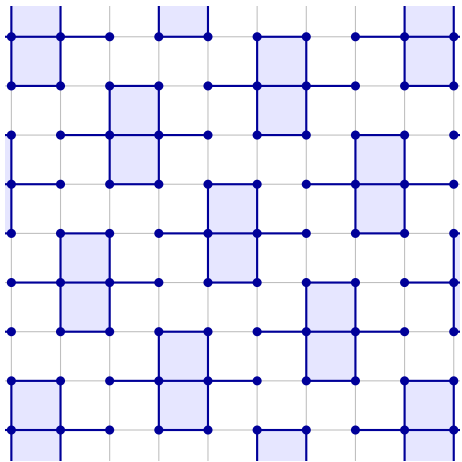}
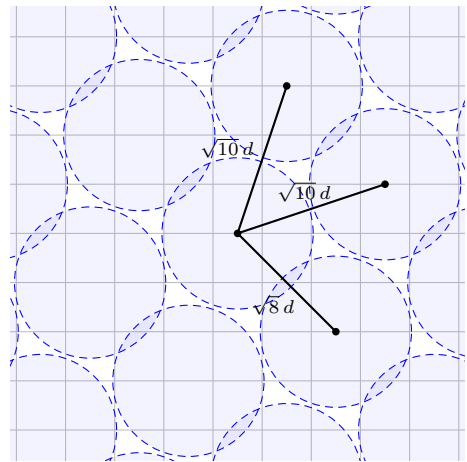
    \caption{8-point motif}
  \end{subfigure}
  \hspace{1em}
  \begin{subfigure}[t]{0.45\linewidth}
    \centering
    \begin{tikzpicture}[scale=1]
  \clip (-2.6750,-3.0000) rectangle (3.3250,3.0000);
  \begin{scope}[lightgray]
    \draw (-5.2000,-4.5500) -- (5.2000,-4.5500);
    \draw (-5.2000,-3.9000) -- (5.2000,-3.9000);
    \draw (-5.2000,-3.2500) -- (5.2000,-3.2500);
    \draw (-5.2000,-2.6000) -- (5.2000,-2.6000);
    \draw (-5.2000,-1.9500) -- (5.2000,-1.9500);
    \draw (-5.2000,-1.3000) -- (5.2000,-1.3000);
    \draw (-5.2000,-0.65000) -- (5.2000,-0.65000);
    \draw (-5.2000,0) -- (5.2000,0);
    \draw (-5.2000,0.65000) -- (5.2000,0.65000);
    \draw (-5.2000,1.3000) -- (5.2000,1.3000);
    \draw (-5.2000,1.9500) -- (5.2000,1.9500);
    \draw (-5.2000,2.6000) -- (5.2000,2.6000);
    \draw (-5.2000,3.2500) -- (5.2000,3.2500);
    \draw (-5.2000,3.9000) -- (5.2000,3.9000);
    \draw (-5.2000,4.5500) -- (5.2000,4.5500);
    \draw (-5.2000,-4.5500) -- (-5.2000,4.5500);
    \draw (-4.5500,-4.5500) -- (-4.5500,4.5500);
    \draw (-3.9000,-4.5500) -- (-3.9000,4.5500);
    \draw (-3.2500,-4.5500) -- (-3.2500,4.5500);
    \draw (-2.6000,-4.5500) -- (-2.6000,4.5500);
    \draw (-1.9500,-4.5500) -- (-1.9500,4.5500);
    \draw (-1.3000,-4.5500) -- (-1.3000,4.5500);
    \draw (-0.65000,-4.5500) -- (-0.65000,4.5500);
    \draw (0,-4.5500) -- (0,4.5500);
    \draw (0.65000,-4.5500) -- (0.65000,4.5500);
    \draw (1.3000,-4.5500) -- (1.3000,4.5500);
    \draw (1.9500,-4.5500) -- (1.9500,4.5500);
    \draw (2.6000,-4.5500) -- (2.6000,4.5500);
    \draw (3.2500,-4.5500) -- (3.2500,4.5500);
    \draw (3.9000,-4.5500) -- (3.9000,4.5500);
    \draw (4.5500,-4.5500) -- (4.5500,4.5500);
    \draw (5.2000,-4.5500) -- (5.2000,4.5500);
  \end{scope}
  \begin{scope}[blue,  densely dashed, fill opacity=0.05]
    \filldraw (-0.97500,-3.9000) circle[radius=1];
    \filldraw (-2.2750,-2.6000) circle[radius=1];
    \filldraw (-3.5750,-1.3000) circle[radius=1];
    \filldraw (0.97500,-3.2500) circle[radius=1];
    \filldraw (-0.32500,-1.9500) circle[radius=1];
    \filldraw (-1.6250,-0.65000) circle[radius=1];
    \filldraw (-2.9250,0.65000) circle[radius=1];
    \filldraw (4.2250,-3.9000) circle[radius=1];
    \filldraw (2.9250,-2.6000) circle[radius=1];
    \filldraw (1.6250,-1.3000) circle[radius=1];
    \filldraw (0.32500,0) circle[radius=1];
    \filldraw (-0.97500,1.3000) circle[radius=1];
    \filldraw (-2.2750,2.6000) circle[radius=1];
    \filldraw (-3.5750,3.9000) circle[radius=1];
    \filldraw (3.5750,-0.65000) circle[radius=1];
    \filldraw (2.2750,0.65000) circle[radius=1];
    \filldraw (0.97500,1.9500) circle[radius=1];
    \filldraw (-0.32500,3.2500) circle[radius=1];
    \filldraw (4.2250,1.3000) circle[radius=1];
    \filldraw (2.9250,2.6000) circle[radius=1];
    \filldraw (1.6250,3.9000) circle[radius=1];
  \end{scope}
  \begin{scope}[black, thick, font=\scriptsize]
    \fill (0.32500,0) circle[radius=0.05];
    \fill (1.6250,-1.3000) circle[radius=0.05];
    \fill (2.2750,0.65000) circle[radius=0.05];
    \fill (0.97500,1.9500) circle[radius=0.05];
    \draw (0.32500,0) -- (1.6250,-1.3000) node[pos=0.73, anchor=east, left=2pt] {$\sqrt8\,d$};
    \draw (0.32500,0) -- (2.2750,0.65000) node[pos=0.45, anchor=south, above=0pt] {$\sqrt{10}\,d$};
    \draw (0.32500,0) -- (0.97500,1.9500) node[pos=0.58, anchor=east, left=1pt] {$\sqrt{10}\,d$};
  \end{scope}
\end{tikzpicture}
    \caption{Overlapping realization}
  \end{subfigure}
  \caption{The invalid motif construction of Alm et~al.\ for the square lattice.}
  \label{fig:square-alm-invalid}
\end{figure}

The invalid construction of Alm et~al.\ is the $8$-point single-family pattern, in which each motif is responsible for $8$ lattice points. It was intended to cover the interval
\[
  \frac2{\sqrt{10}} \le d \le \frac23.
\]
Conceptually, each realizing disk has six neighboring disks. By rotational symmetry, only three of these neighbor relations are distinct. The first two give the non-overlap constraint
\[
  d \ge \frac2{\sqrt{10}},
\]
which is the lower endpoint stated in their construction. The third neighbor relation, however, gives the stronger constraint
\[
  d \ge \frac1{\sqrt2}.
\]
On the other hand, the containment condition for the motif requires
\[
  d \le \frac23.
\]
Since $1/\sqrt2 > 2/3$, no spacing $d$ satisfies both constraints. Thus the proposed realization does not yield a valid motif pattern. The obstruction is illustrated in \cref{fig:square-alm-invalid}.

\subsection{A mixed-motif recovery}
\label{sec:square-mixed}

\begin{figure}[!t]
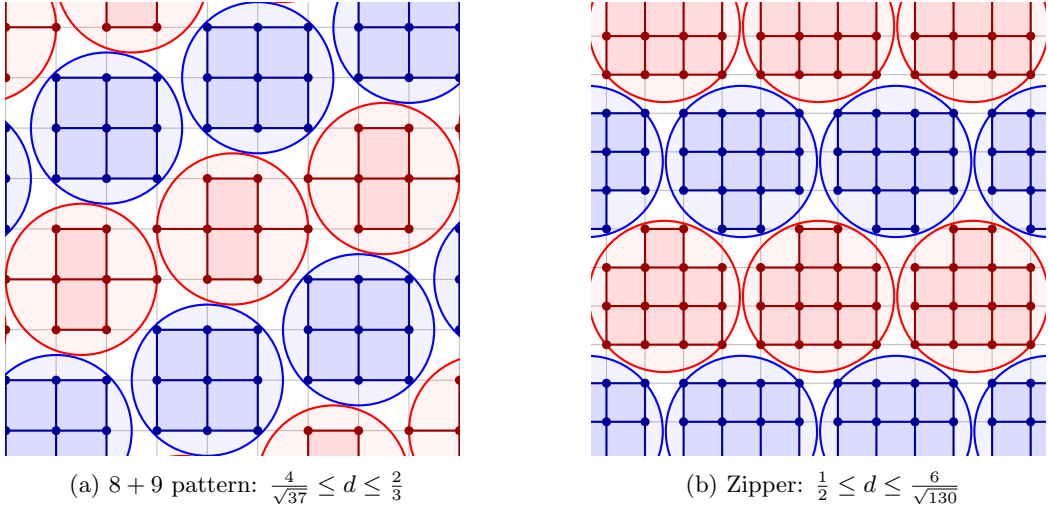

  \centering
  \begin{subfigure}[t]{0.45\linewidth}
    \centering
    \input{img/square/mono8mono9.tikz}
    \caption{$8+9$ pattern: $\frac4{\sqrt{37}} \le d \le \frac23$}
    \label{fig:square-mono8mono9}
  \end{subfigure}
  \hspace{1em}
  \begin{subfigure}[t]{0.45\linewidth}
    \centering
    \input{img/square/duo14.tikz}
    \caption{Zipper: $\frac12 \le d \le \frac6{\sqrt{130}}$}
    \label{fig:square-duo14}
  \end{subfigure}
  \caption{The two new motif patterns used for the square lattice.}
  \label{fig:square-new-patterns}
\end{figure}

We now modify the invalid construction to recover part of the missing interval. Although disks in neighboring rows of the original pattern overlap, neighboring disks within each row do not. We therefore begin by retaining a single row of $8$-point motifs and discarding all other rows. We place beside it a row of big-square motifs from \cref{fig:square-uniform9}, whose disks are responsible for $9$ lattice points each. These two rows form a fundamental strip, which we repeat periodically to obtain the two-family motif pattern shown in \cref{fig:square-mono8mono9}.

Equivalently, one may view the construction as deleting alternating rows from the original $8$-point pattern, but only after increasing the separation between the surviving rows. Merely deleting alternating rows leaves insufficient space for the intervening $9$-point rows; the enlarged row separation is therefore an essential part of the construction.

The containment condition for the $8$-point family is the same as in the original construction, while the containment condition for the $9$-point family is that of the big-square construction. The new constraint comes from non-overlap between disks belonging to different families. The minimum distance between centers of disks in different families gives
\[
  d \ge \frac4{\sqrt{37}}.
\]
The upper endpoint is inherited from the $8$-point family:
\[
  d \le \frac23.
\]
Therefore, this two-family motif pattern is valid for
\[
  \frac4{\sqrt{37}} \le d \le \frac23.
\]

\subsection{Zipper pattern}
\label{sec:square-zipper}

We next give a new construction for smaller spacings. The motif is a $14$-point ``zipper'' shape obtained from a $4\times4$ block of square-lattice points by deleting two adjacent corner points on one side. We arrange copies of this motif in rows. In alternating rows, we use the reflected motif, shifted so that the two types of rows interlock; see \cref{fig:square-duo14}. This gives a two-family motif pattern, with each family responsible for $14$ points per motif.

It is straightforward to verify from the row construction that the motif copies tile the square lattice. The main issue is to choose realizing disk centers so that every motif is covered and no two realizing disks overlap. We use a piecewise choice of centers, depending on the spacing $d$.

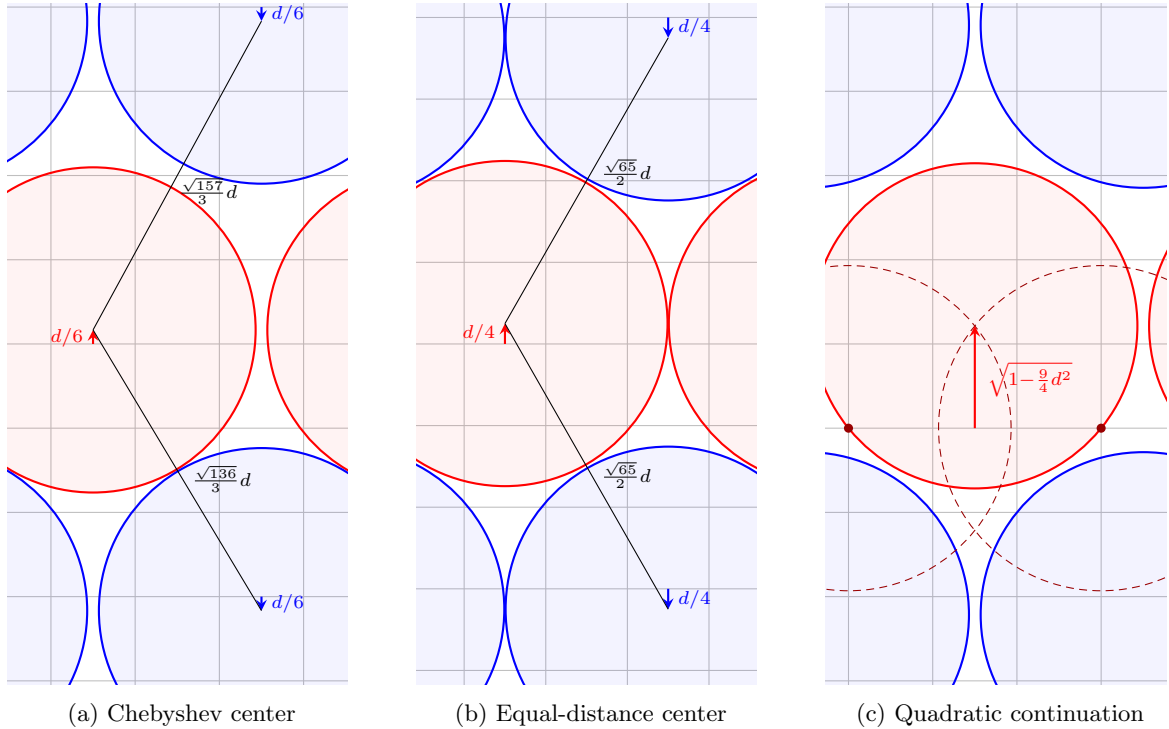
\begin{figure}[!b]
  \centering
  \begin{subfigure}[t]{0.32\linewidth}
    \centering
    \begin{tikzpicture}[scale=2.15]
  \clip (-0.26951,-2.0930) rectangle (1.8235,2.0930);
  \begin{scope}[lightgray]
    \draw (-3.1080,-3.6260) -- (3.1080,-3.6260);
    \draw (-3.1080,-3.1080) -- (3.1080,-3.1080);
    \draw (-3.1080,-2.5900) -- (3.1080,-2.5900);
    \draw (-3.1080,-2.0720) -- (3.1080,-2.0720);
    \draw (-3.1080,-1.5540) -- (3.1080,-1.5540);
    \draw (-3.1080,-1.0360) -- (3.1080,-1.0360);
    \draw (-3.1080,-0.51800) -- (3.1080,-0.51800);
    \draw (-3.1080,0) -- (3.1080,0);
    \draw (-3.1080,0.51800) -- (3.1080,0.51800);
    \draw (-3.1080,1.0360) -- (3.1080,1.0360);
    \draw (-3.1080,1.5540) -- (3.1080,1.5540);
    \draw (-3.1080,2.0720) -- (3.1080,2.0720);
    \draw (-3.1080,2.5900) -- (3.1080,2.5900);
    \draw (-3.1080,3.1080) -- (3.1080,3.1080);
    \draw (-3.1080,3.6260) -- (3.1080,3.6260);
    \draw (-3.1080,-3.6260) -- (-3.1080,3.6260);
    \draw (-2.5900,-3.6260) -- (-2.5900,3.6260);
    \draw (-2.0720,-3.6260) -- (-2.0720,3.6260);
    \draw (-1.5540,-3.6260) -- (-1.5540,3.6260);
    \draw (-1.0360,-3.6260) -- (-1.0360,3.6260);
    \draw (-0.51800,-3.6260) -- (-0.51800,3.6260);
    \draw (0,-3.6260) -- (0,3.6260);
    \draw (0.51800,-3.6260) -- (0.51800,3.6260);
    \draw (1.0360,-3.6260) -- (1.0360,3.6260);
    \draw (1.5540,-3.6260) -- (1.5540,3.6260);
    \draw (2.0720,-3.6260) -- (2.0720,3.6260);
    \draw (2.5900,-3.6260) -- (2.5900,3.6260);
    \draw (3.1080,-3.6260) -- (3.1080,3.6260);
  \end{scope}
  \begin{scope}[red, thick, fill opacity=0.05]
    \filldraw (0.25900,0.086333) circle[radius=1];
    \filldraw (2.3310,0.086333) circle[radius=1];
  \end{scope}
  \begin{scope}[blue, thick, fill opacity=0.05]
    \filldraw (-0.77700,-1.6403) circle[radius=1];
    \filldraw (-0.77700,1.9857) circle[radius=1];
    \filldraw (1.2950,-1.6403) circle[radius=1];
    \filldraw (1.2950,1.9857) circle[radius=1];
  \end{scope}
  \begin{scope}[-stealth, thick, font=\scriptsize]
    \draw[red] (0.25900,0) -- (0.25900,0.086333) node[midway, anchor=east] {$d/6$};
    \draw[blue] (1.2950,-1.5540) -- (1.2950,-1.6403) node[midway, anchor=west] {$d/6$};
    \draw[blue] (1.2950,2.0720) -- (1.2950,1.9857) node[midway, anchor=west] {$d/6$};
  \end{scope}
  \begin{scope}[black, font=\scriptsize]
    \draw (0.25900,0.086333) -- (1.2950,-1.6403) node[pos=0.53, anchor=west] {$\frac{\sqrt{136}}{3}d$};
    \draw (0.25900,0.086333) -- (1.2950,1.9857) node[pos=0.45, anchor=west] {$\frac{\sqrt{157}}{3}d$};
  \end{scope}
\end{tikzpicture}
    \caption{Chebyshev center}
    \label{fig:square-duo14-chebyshev}
  \end{subfigure}
  \hfill
  \begin{subfigure}[t]{0.32\linewidth}
    \centering
    \begin{tikzpicture}[scale=2.15]
  \clip (-0.29351,-2.0930) rectangle (1.7995,2.0930);
  \begin{scope}[lightgray]
    \draw (-3.0120,-3.5140) -- (3.0120,-3.5140);
    \draw (-3.0120,-3.0120) -- (3.0120,-3.0120);
    \draw (-3.0120,-2.5100) -- (3.0120,-2.5100);
    \draw (-3.0120,-2.0080) -- (3.0120,-2.0080);
    \draw (-3.0120,-1.5060) -- (3.0120,-1.5060);
    \draw (-3.0120,-1.0040) -- (3.0120,-1.0040);
    \draw (-3.0120,-0.50200) -- (3.0120,-0.50200);
    \draw (-3.0120,0) -- (3.0120,0);
    \draw (-3.0120,0.50200) -- (3.0120,0.50200);
    \draw (-3.0120,1.0040) -- (3.0120,1.0040);
    \draw (-3.0120,1.5060) -- (3.0120,1.5060);
    \draw (-3.0120,2.0080) -- (3.0120,2.0080);
    \draw (-3.0120,2.5100) -- (3.0120,2.5100);
    \draw (-3.0120,3.0120) -- (3.0120,3.0120);
    \draw (-3.0120,3.5140) -- (3.0120,3.5140);
    \draw (-3.0120,-3.5140) -- (-3.0120,3.5140);
    \draw (-2.5100,-3.5140) -- (-2.5100,3.5140);
    \draw (-2.0080,-3.5140) -- (-2.0080,3.5140);
    \draw (-1.5060,-3.5140) -- (-1.5060,3.5140);
    \draw (-1.0040,-3.5140) -- (-1.0040,3.5140);
    \draw (-0.50200,-3.5140) -- (-0.50200,3.5140);
    \draw (0,-3.5140) -- (0,3.5140);
    \draw (0.50200,-3.5140) -- (0.50200,3.5140);
    \draw (1.0040,-3.5140) -- (1.0040,3.5140);
    \draw (1.5060,-3.5140) -- (1.5060,3.5140);
    \draw (2.0080,-3.5140) -- (2.0080,3.5140);
    \draw (2.5100,-3.5140) -- (2.5100,3.5140);
    \draw (3.0120,-3.5140) -- (3.0120,3.5140);
  \end{scope}
  \begin{scope}[red, thick, fill opacity=0.05]
    \filldraw (0.25100,0.12550) circle[radius=1];
    \filldraw (2.2590,0.12550) circle[radius=1];
  \end{scope}
  \begin{scope}[blue, thick, fill opacity=0.05]
    \filldraw (-0.75300,-1.6315) circle[radius=1];
    \filldraw (-0.75300,1.8825) circle[radius=1];
    \filldraw (1.2550,-1.6315) circle[radius=1];
    \filldraw (1.2550,1.8825) circle[radius=1];
  \end{scope}
  \begin{scope}[-stealth, thick, font=\scriptsize]
    \draw[red] (0.25100,0) -- (0.25100,0.12550) node[midway, anchor=east] {$d/4$};
    \draw[blue] (1.2550,-1.5060) -- (1.2550,-1.6315) node[midway, anchor=west] {$d/4$};
    \draw[blue] (1.2550,2.0080) -- (1.2550,1.8825) node[midway, anchor=west] {$d/4$};
  \end{scope}
  \begin{scope}[black, font=\scriptsize]
    \draw (0.25100,0.12550) -- (1.2550,-1.6315) node[pos=0.53, anchor=west] {$\frac{\sqrt{65}}{2}d$};
    \draw (0.25100,0.12550) -- (1.2550,1.8825) node[pos=0.53, anchor=west] {$\frac{\sqrt{65}}{2}d$};
  \end{scope}
\end{tikzpicture}
    \caption{Equal-distance center}
    \label{fig:square-duo14-equaldist}
  \end{subfigure}
  \hfill
  \begin{subfigure}[t]{0.32\linewidth}
    \centering
    \begin{tikzpicture}[scale=2.15]
  \clip (-0.65801,-2.0930) rectangle (1.4350,2.0930);
  \begin{scope}[lightgray]
    \draw (-3.1080,-3.6260) -- (3.1080,-3.6260);
    \draw (-3.1080,-3.1080) -- (3.1080,-3.1080);
    \draw (-3.1080,-2.5900) -- (3.1080,-2.5900);
    \draw (-3.1080,-2.0720) -- (3.1080,-2.0720);
    \draw (-3.1080,-1.5540) -- (3.1080,-1.5540);
    \draw (-3.1080,-1.0360) -- (3.1080,-1.0360);
    \draw (-3.1080,-0.51800) -- (3.1080,-0.51800);
    \draw (-3.1080,0) -- (3.1080,0);
    \draw (-3.1080,0.51800) -- (3.1080,0.51800);
    \draw (-3.1080,1.0360) -- (3.1080,1.0360);
    \draw (-3.1080,1.5540) -- (3.1080,1.5540);
    \draw (-3.1080,2.0720) -- (3.1080,2.0720);
    \draw (-3.1080,2.5900) -- (3.1080,2.5900);
    \draw (-3.1080,3.1080) -- (3.1080,3.1080);
    \draw (-3.1080,3.6260) -- (3.1080,3.6260);
    \draw (-3.1080,-3.6260) -- (-3.1080,3.6260);
    \draw (-2.5900,-3.6260) -- (-2.5900,3.6260);
    \draw (-2.0720,-3.6260) -- (-2.0720,3.6260);
    \draw (-1.5540,-3.6260) -- (-1.5540,3.6260);
    \draw (-1.0360,-3.6260) -- (-1.0360,3.6260);
    \draw (-0.51800,-3.6260) -- (-0.51800,3.6260);
    \draw (0,-3.6260) -- (0,3.6260);
    \draw (0.51800,-3.6260) -- (0.51800,3.6260);
    \draw (1.0360,-3.6260) -- (1.0360,3.6260);
    \draw (1.5540,-3.6260) -- (1.5540,3.6260);
    \draw (2.0720,-3.6260) -- (2.0720,3.6260);
    \draw (2.5900,-3.6260) -- (2.5900,3.6260);
    \draw (3.1080,-3.6260) -- (3.1080,3.6260);
  \end{scope}
  \begin{scope}[red, thick, fill opacity=0.05]
    \filldraw (0.25900,0.11150) circle[radius=1];
    \filldraw (2.3310,0.11150) circle[radius=1];
  \end{scope}
  \begin{scope}[blue, thick, fill opacity=0.05]
    \filldraw (-0.77700,-1.6655) circle[radius=1];
    \filldraw (-0.77700,1.9605) circle[radius=1];
    \filldraw (1.2950,-1.6655) circle[radius=1];
    \filldraw (1.2950,1.9605) circle[radius=1];
  \end{scope}
  \begin{scope}[red!60!black, densely dashed]
    \fill (-0.51800,-0.51800) circle[radius=0.027907];
    \fill (1.0360,-0.51800) circle[radius=0.027907];
    \draw (-0.51800,-0.51800) circle[radius=1];
    \draw (1.0360,-0.51800) circle[radius=1];
  \end{scope}
  \begin{scope}[-stealth, thick, font=\scriptsize]
    \draw[red] (0.25900,-0.51800) -- (0.25900,0.11150) node[pos=0.48, anchor=west, right=1pt] {$\sqrt{1{-}\frac94d^2}$};
  \end{scope}
\end{tikzpicture}
    \caption{Quadratic continuation}
    \label{fig:square-duo14-quadratic}
  \end{subfigure}
  \caption{Three choices of realizing centers for the zipper pattern.}
  \label{fig:square-duo14-center}
\end{figure}

A first natural choice is the Chebyshev center of the $14$-point motif; see \cref{fig:square-duo14-chebyshev}. With this center, the farthest assigned lattice points lie at distance
\[
  \frac{\sqrt{130}}{6}d.
\]
Thus the containment condition gives
\[
  d \le \frac6{\sqrt{130}}.
\]
This determines the largest spacing attainable by this motif. However, the Chebyshev centers are not optimal for non-overlap. As $d$ decreases, disks from different families approach each other in one direction while remaining farther apart in another direction.

For smaller spacings, we therefore move the centers so that the two relevant inter-family distances are equal; see \cref{fig:square-duo14-equaldist}. With this choice, the first kissing constraint occurs between disks belonging to the same family. This gives the lower bound
\[
  d \ge \frac12.
\]

The containment condition for these equal-distance centers holds up to
\[
  d \le \frac4{\sqrt{61}},
\]
at which point the non-deleted corner points of the motif lie on the boundary of the disk.

For larger spacings, we use a one-parameter family of centers. Geometrically, the center is chosen as the intersection of two unit circles centered at two non-deleted corner points of the motif. Equivalently, the center is shifted from the baseline joining these two points by
\[
  \sqrt{1 - \left(\frac{3}{2}d\right)^2},
\]
as illustrated in \cref{fig:square-duo14-quadratic}. This quadratic continuation is valid for
\[
  \frac{24}{\sqrt{2257}} \le d \le \frac6{\sqrt{130}}.
\]
Since
\[
  \frac{24}{\sqrt{2257}} < \frac{4}{\sqrt{61}},
\]
we may start the quadratic continuation at the equal-distance center when $d=4/\sqrt{61}$. As $d$ increases, this family of centers varies continuously and reaches the Chebyshev center when $d=6/\sqrt{130}$. Its non-overlap constraints are inactive throughout the interval
\[
  \frac4{\sqrt{61}} \le d \le \frac6{\sqrt{130}}
\]
and hence it fills the gap between the two simpler center choices.

Combining the equal-distance centers with this quadratic continuation gives a valid realization of the zipper pattern for
\[
  \frac12 \le d \le \frac6{\sqrt{130}}.
\]
Together with the patterns discussed above, this proves \cref{thm:square-spacing}.

\section{Discussion and Open Questions}
\label{sec:discuss}

We have constructed periodic coverings of the triangular lattice, the honeycomb point set, and the square lattice by pairwise non-overlapping unit disks. For the honeycomb point set, native pair, hexagon, and 24-point motifs simplify constructions inherited from the triangular lattice and close two gaps between inherited intervals. For the square lattice, we identified an unintended overlap in a construction of Alm et~al.~\cite{alm2013motif} and introduced two patterns that recover part of the affected interval and establish an additional interval at smaller spacings. Nevertheless, the intervals proved here are constructive sufficient conditions, not complete characterizations of coverability.

Most of our triangular-lattice constructions follow a common geometric framework: at their lower endpoints, the realizing disks form a hexagonal packing. Within this framework, the density calculation in \cref{rmk:tri-density} predicts the lower endpoint directly from the average number of lattice points assigned to each disk. This principle substantially reduces the search space. Once one translation vector for the center lattice is chosen, the hexagonal symmetry determines a compatible second vector, so candidate patterns can often be described using essentially one independent translation direction. We developed a computer-assisted search around this framework, through which many of the patterns reported in \cref{sec:tri} were discovered.

The same structure also indicates why several gaps in \cref{thm:tri-spacing} may be difficult to close using further hexagonal-packing patterns. Consider, for example, the gap
\[
  \frac5{\sqrt{91}} < d < \frac2{\sqrt{13}}.
\]
For a single-family hexagonal-packing construction whose disks are each responsible for $k$ points, \cref{rmk:tri-density} gives the lower endpoint $2/\sqrt{k}$. A 14-point motif could narrow this gap, while a 15-point motif would have a sufficiently small density bound to meet the interval on its lower side, provided that its containment constraint extends far enough. Such motifs are not natural among the symmetric vertex-centered neighborhoods of the triangular lattice: the 13-point neighborhood is obtained by centering a disk at a lattice point, but the next radial shell contains six points, increasing the responsibility directly from $13$ to $19$. Similar shell effects arise from the sixfold symmetry of the lattice.

This observation is a limitation of the framework, not an impossibility result. A covering need not have hexagonally packed centers, nor must its motifs be symmetric neighborhoods of a vertex or face. The alternating-pair pattern in \cref{fig:tri-duo2} and the cyclone pattern in \cref{fig:tri-cyclone} already demonstrate that useful constructions can depart from the hexagonal-packing structure. Less symmetric motifs, non-triangular center sets, or patterns with several interacting families may therefore close gaps that appear inaccessible to the density-guided search.

The contrast with the square lattice further clarifies the role of symmetry. For the triangular lattice, the hexagonal-packing framework relates the two generators of the center lattice by a $60^\circ$ rotation. A square-lattice motif pattern generally has no such reduction: its rectangular periodic structure requires two independent translation vectors. Consequently, a comparable computer search must consider a larger parameter space, including two period directions and the relative placement of multiple motif families. The zipper construction illustrates the additional flexibility that may be required, since even a fixed motif pattern uses different realizing centers over different subintervals of $d$.

At the opposite end of the spacing range, our smallest triangular-lattice construction occurs at
\[
  d=\frac2{\sqrt{37}}\approx0.329.
\]
A simple geometric obstruction suggests a smaller natural scale. Three mutually tangent unit disks leave a curvilinear triangular gap whose inscribed radius is $2/\sqrt3-1$, whereas the covering radius of $\mathcal T_d$ is $d/\sqrt3$. Comparing these quantities gives
\[
  d\le2-\sqrt3\approx0.268.
\]
Below this scale, every translate of the largest gap in a hexagonal packing contains a point of $\mathcal T_d$. The separation between this obstruction and our smallest construction leaves substantial room for either stronger impossibility arguments or denser constructions with a different geometry.

These observations lead to the following open questions.

\begin{enumerate}
  \item For each of the three point sets considered here, what is the complete set of spacings that admit a covering by pairwise non-overlapping unit disks?

  \item Which of the bounded gaps in \cref{thm:tri-spacing} can be narrowed or closed by patterns that do not exhibit a hexagonal-packing structure?

  \item Is the scale $d=2-\sqrt3$ related to the true lower threshold for the triangular lattice, or is the threshold strictly larger?

  \item Can the computer-assisted search be extended to enumerate and verify less symmetric motif patterns, including patterns with non-triangular center sets and two independent translation vectors?
\end{enumerate}

\bibliographystyle{abbrv}
\bibliography{ref}

\appendix

\section{Details of the cyclone pattern}
\label{app:tri-cyclone}

The cyclone pattern, shown in \cref{fig:tri-cyclone}, is a $25$-family periodic pattern. It has a valid realization for
\[
  d \in \left[ \frac{15+\sqrt{85}}{14} , 2 \right].
\]
Since
\[
  \frac{15+\sqrt{85}}{14} < \sqrt{3},
\]
this pattern covers the remaining gap
\[
  \sqrt{3} < d < 2
\]
between the primitive singleton and triangle patterns from \cref{fig:tri-uniform1,fig:tri-uniform3}.

Although the pattern has $25$ disk families, its construction is most easily described in terms of two component types:
\begin{itemize}
  \item a hexagon component, consisting of $7$ one-point motifs; and
  \item a bent component, consisting of $3$ two-point motifs.
\end{itemize}
The bent component has six rotational variants, obtained by rotations through multiples of $60^\circ$. We call a bent component \emph{even} if it is obtained by a rotation through $0^\circ$, $120^\circ$, or $240^\circ$, and \emph{odd} otherwise.

A fundamental tile of the cyclone pattern consists of one hexagon component and six bent components, one in each rotational variant. Thus, each fundamental tile is responsible for
\[
  7 + 6{\times}6 = 43
\]
lattice points.

\begin{figure}[!b]
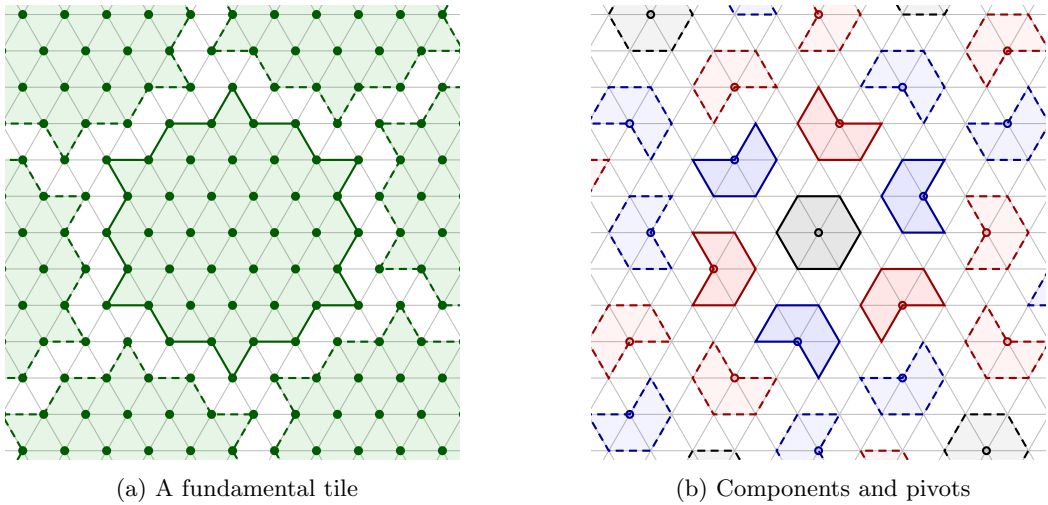

  \centering
  \begin{subfigure}[t]{0.45\linewidth}
    \centering
    \input{img/tri/cyclone-motif-fundamental.tikz}
    \caption{A fundamental tile}
    \label{fig:tri-cyclone-motif-fundamental}
  \end{subfigure}
  \hspace{1em}
  \begin{subfigure}[t]{0.45\linewidth}
    \centering
    \input{img/tri/cyclone-motif-component.tikz}
    \caption{Components and pivots}
    \label{fig:tri-cyclone-motif-component}
  \end{subfigure}
  \caption{The motif structure of the cyclone pattern.}
  \label{fig:tri-cyclone-motif}
\end{figure}

Strictly speaking, a motif is a set of lattice points assigned to a single disk. In this appendix, we use a \emph{component} as a convenient shorthand for a small collection of motifs that will be repeated together. Thus \cref{fig:tri-cyclone-motif-fundamental} shows the motifs in one full tile of the pattern, while \cref{fig:tri-cyclone-motif-component} shows their grouping into components and marks the corresponding pivot points.

The construction data are listed in \cref{tab:tri-cyclone}. For a row with pivot $p$, direction vector $u$, and offset length $\ell$, the disk center is
\[
  dp + \ell\,\frac{u}{|u|}.
\]
Here the direction vectors are written in Eisenstein coordinates. Thus the direction column specifies only the direction of the offset; the vector is normalized before being multiplied by the listed length. The central disk of the hexagon component has zero offset.

\begin{table}[t]
  \centering
  \begin{tabular}{crccc}
    \toprule
    \multicolumn{2}{c}{\multirow{2}{*}{Family}}
    & \multirow{2}{*}{Pivot}
    & \multicolumn{2}{c}{Offset} \\
    \cmidrule(lr){4-5}
    &&& Direction & Length \\
    \midrule
    \multirow{7}{*}{\rotatebox{90}{hexagon}}
    & 1  & $0$ &        none &     $0$ \\
    & 2  & $0$ &         $1$ & $2$ \\
    & 3  & $0$ &  $1+\omega$ & $2$ \\
    & 4  & $0$ &    $\omega$ & $2$ \\
    & 5  & $0$ &        $-1$ & $2$ \\
    & 6  & $0$ & $-1-\omega$ & $2$ \\
    & 7  & $0$ &   $-\omega$ & $2$ \\
    \midrule
    \multirow{9}{*}{\rotatebox{90}{even bents}}
    & 8  &   $3+\omega$ &         $-1$ &      $1$ \\
    & 9  &   $3+\omega$ & $-1-2\omega$ & $\sqrt3$ \\
    & 10 &   $3+\omega$ &  $1+2\omega$ & $\sqrt3$ \\
    & 11 & $-1+2\omega$ &    $-\omega$ &      $1$ \\
    & 12 & $-1+2\omega$ &   $2+\omega$ & $\sqrt3$ \\
    & 13 & $-1+2\omega$ &  $-2-\omega$ & $\sqrt3$ \\
    & 14 & $-2-3\omega$ &   $1+\omega$ &      $1$ \\
    & 15 & $-2-3\omega$ &   $1-\omega$ & $\sqrt3$ \\
    & 16 & $-2-3\omega$ &  $-1+\omega$ & $\sqrt3$ \\
    \midrule
    \multirow{9}{*}{\rotatebox{90}{odd bents}}
    & 17 &  $-3-\omega$ &          $1$ &      $1$ \\
    & 18 &  $-3-\omega$ &  $1+2\omega$ & $\sqrt3$ \\
    & 19 &  $-3-\omega$ & $-1-2\omega$ & $\sqrt3$ \\
    & 20 &  $1-2\omega$ &     $\omega$ &      $1$ \\
    & 21 &  $1-2\omega$ &  $-2-\omega$ & $\sqrt3$ \\
    & 22 &  $1-2\omega$ &   $2+\omega$ & $\sqrt3$ \\
    & 23 &  $2+3\omega$ &  $-1-\omega$ &      $1$ \\
    & 24 &  $2+3\omega$ &  $-1+\omega$ & $\sqrt3$ \\
    & 25 &  $2+3\omega$ &   $1-\omega$ & $\sqrt3$ \\
    \bottomrule
  \end{tabular}
  \caption{Construction data for the disk families in the cyclone pattern.}
  \label{tab:tri-cyclone}
\end{table}

\subsection{Hexagon component}

The hexagon component consists of seven disks arranged as in the hexagonal packing. One disk is centered at the pivot lattice point, and the other six disks kiss it in the six lattice directions. Each disk is responsible for one lattice point.

The central disk realizes its one-point motif whenever no adjacent lattice point lies in its interior. This requires
\[
  d \ge 1.
\]
For each surrounding disk, the assigned lattice point lies on the corresponding radial lattice direction. The disk covers this point exactly when
\[
  1 \le d \le 3.
\]
Hence the hexagon component is valid for
\[
  1 \le d \le 3.
\]

\subsection{Bent component}

A bent component consists of three disks. Each disk is responsible for two lattice points, so the component as a whole is responsible for six points. Without loss of generality, these six points are
\[
  \{0,-1,1+\omega,\omega,-\omega,-1-\omega\}.
\]
The three disk centers form a partial hexagonal-packing configuration: the two outer disks kiss the middle disk, and the angle between the two center-to-center directions is $120^\circ$.

By reflection symmetry, it is enough to check the middle disk and one outer disk. The middle disk is responsible for $0$ and $-1$. The point $0$ is covered by construction, and the point $-1$ is covered exactly in the range
\[
  1 \le d \le 2.
\]
For an outer disk, we may take the assigned points to be $\omega$ and $1+\omega$. These two points are covered in the same range,
\[
  1 \le d \le 2.
\]
Therefore each bent component is valid for
\[
  1 \le d \le 2.
\]

\subsection{Interactions between components}

It remains to verify the non-overlap between disks belonging to different components. There are three types of possible cross-component contacts:
\begin{enumerate}
  \item contacts between bent components of the same parity;
  \item contacts between bent components of opposite parity; and
  \item contacts between a bent component and a hexagon component.
\end{enumerate}
Let $\ell_1$, $\ell_2$, and $\ell_3$ denote the corresponding distances between disk centers in these three cases. As shown in \cref{fig:tri-cyclone-kissing}, each case gives the same squared distance:
\[
  \ell_t^2 = 7d^2 - 15d + 9,
  \qquad t\in\{1,2,3\}.
\]
Therefore, non-overlap is equivalent to requiring $\ell_t\ge 2$ for each $t$, which yields
\[
  d \ge \frac{15+\sqrt{85}}{14}.
\]

Combining this cross-component constraint with the component-wise validity ranges gives a valid realization of the cyclone pattern for
\[
  \frac{15+\sqrt{85}}{14} \le d \le 2.
\]

\begin{figure}[!t]
  \centering
  \begin{tikzpicture}[scale=1.25]
  \clip (-1.3076,-3.5491) rectangle (9.0924,2.0509);
  \begin{scope}[lightgray]
    \draw (-13.840,-7.4910) -- (13.840,-7.4910);
    \draw (-13.840,-5.9928) -- (13.840,-5.9928);
    \draw (-13.840,-4.4946) -- (13.840,-4.4946);
    \draw (-13.840,-2.9964) -- (13.840,-2.9964);
    \draw (-13.840,-1.4982) -- (13.840,-1.4982);
    \draw (-13.840,0) -- (13.840,0);
    \draw (-13.840,1.4982) -- (13.840,1.4982);
    \draw (-13.840,2.9964) -- (13.840,2.9964);
    \draw (-13.840,4.4946) -- (13.840,4.4946);
    \draw (-13.840,5.9928) -- (13.840,5.9928);
    \draw (-13.840,7.4910) -- (13.840,7.4910);
    \draw (-21.625,-7.4910) -- (-12.975,7.4910);
    \draw (-19.895,-7.4910) -- (-11.245,7.4910);
    \draw (-18.165,-7.4910) -- (-9.5148,7.4910);
    \draw (-16.435,-7.4910) -- (-7.7849,7.4910);
    \draw (-14.705,-7.4910) -- (-6.0549,7.4910);
    \draw (-12.975,-7.4910) -- (-4.3249,7.4910);
    \draw (-11.245,-7.4910) -- (-2.5950,7.4910);
    \draw (-9.5148,-7.4910) -- (-0.86498,7.4910);
    \draw (-7.7849,-7.4910) -- (0.86498,7.4910);
    \draw (-6.0549,-7.4910) -- (2.5950,7.4910);
    \draw (-4.3249,-7.4910) -- (4.3249,7.4910);
    \draw (-2.5950,-7.4910) -- (6.0549,7.4910);
    \draw (-0.86498,-7.4910) -- (7.7849,7.4910);
    \draw (0.86498,-7.4910) -- (9.5148,7.4910);
    \draw (2.5950,-7.4910) -- (11.245,7.4910);
    \draw (4.3249,-7.4910) -- (12.975,7.4910);
    \draw (6.0549,-7.4910) -- (14.705,7.4910);
    \draw (7.7849,-7.4910) -- (16.435,7.4910);
    \draw (9.5148,-7.4910) -- (18.165,7.4910);
    \draw (11.245,-7.4910) -- (19.895,7.4910);
    \draw (12.975,-7.4910) -- (21.625,7.4910);
    \draw (-12.975,-7.4910) -- (-21.625,7.4910);
    \draw (-11.245,-7.4910) -- (-19.895,7.4910);
    \draw (-9.5148,-7.4910) -- (-18.165,7.4910);
    \draw (-7.7849,-7.4910) -- (-16.435,7.4910);
    \draw (-6.0549,-7.4910) -- (-14.705,7.4910);
    \draw (-4.3249,-7.4910) -- (-12.975,7.4910);
    \draw (-2.5950,-7.4910) -- (-11.245,7.4910);
    \draw (-0.86498,-7.4910) -- (-9.5148,7.4910);
    \draw (0.86498,-7.4910) -- (-7.7849,7.4910);
    \draw (2.5950,-7.4910) -- (-6.0549,7.4910);
    \draw (4.3249,-7.4910) -- (-4.3249,7.4910);
    \draw (6.0549,-7.4910) -- (-2.5950,7.4910);
    \draw (7.7849,-7.4910) -- (-0.86498,7.4910);
    \draw (9.5148,-7.4910) -- (0.86498,7.4910);
    \draw (11.245,-7.4910) -- (2.5950,7.4910);
    \draw (12.975,-7.4910) -- (4.3249,7.4910);
    \draw (14.705,-7.4910) -- (6.0549,7.4910);
    \draw (16.435,-7.4910) -- (7.7849,7.4910);
    \draw (18.165,-7.4910) -- (9.5148,7.4910);
    \draw (19.895,-7.4910) -- (11.245,7.4910);
    \draw (21.625,-7.4910) -- (12.975,7.4910);
  \end{scope}
  \begin{scope}[black, densely dashed, fill opacity=0.05]
    \filldraw (0,0) circle[radius=1];
    \filldraw (2.0000,0) circle[radius=1];
    \filldraw (-2.0000,0) circle[radius=1];
    \filldraw (1.0000,1.7320) circle[radius=1];
    \filldraw (-1.0000,1.7320) circle[radius=1];
    \filldraw (1.0000,-1.7320) circle[radius=1];
    \filldraw (-1.0000,-1.7320) circle[radius=1];
    \filldraw (9.2448,1.4982) circle[radius=1];
  \end{scope}
  \begin{scope}[blue, densely dashed, fill opacity=0.05]
    \filldraw (-0.36498,-3.6286) circle[radius=1];
    \filldraw (3.3249,1.4982) circle[radius=1];
    \filldraw (4.3249,-0.23386) circle[radius=1];
    \filldraw (8.8798,-2.1304) circle[radius=1];
  \end{scope}
  \begin{scope}[red, densely dashed, fill opacity=0.05]
    \filldraw (7.9199,0) circle[radius=1];
    \filldraw (6.9199,-1.7320) circle[radius=1];
    \filldraw (6.9199,1.7320) circle[radius=1];
    \filldraw (2.9599,-2.1304) circle[radius=1];
    \filldraw (4.9599,-2.1304) circle[radius=1];
    \filldraw (1.9599,-3.8624) circle[radius=1];
    \filldraw (6.2849,-3.6286) circle[radius=1];
  \end{scope}
  \begin{scope}[purple!80!black, thick, font=\scriptsize]
    \draw (2,0)
       -- (3.3249,0) node[midway, anchor=north]
          {$\frac52d{-}3$}
       -- (3.3249,1.4982) node[midway, anchor=west]
          {$\frac{\sqrt3}2d$};
    \draw (3.2249,0) -- (3.2249,0.1) -- (3.3249,0.1);
    \draw (4.3249,-0.23386)
       -- (4.9599,-0.23386) node[midway, anchor=south]
          {$\frac{3-d}2$}
       -- (4.9599,-2.1304) node[pos=0.4, anchor=west]
          {$\frac{\sqrt{27}}2(d{-}1)$};
    \draw (4.8599,-0.23386) -- (4.8599,-0.33386) -- (4.9599,-0.33386);
    \draw (6.9199,-1.7320)
       -- (6.9199,-2.1304) node[midway, anchor=west]
          {$\sqrt3(d{-}\frac32)$}
       -- (4.9599,-2.1304) node[pos=0.44, anchor=north]
          {$2d{-}\frac32$};
    \draw (6.8199,-2.1304) -- (6.8199,-2.0304) -- (6.9199,-2.0304);
  \end{scope}
  \begin{scope}[black, thick]
    \fill (0,0) circle[radius=0.05];
    \fill (2,0) circle[radius=0.05];
    \draw (0,0) -- (2,0) -- (2.6624,0.7490);
  \end{scope}
  \begin{scope}[blue, thick]
    \fill (4.3249,1.4982) circle[radius=0.05];
    \fill (3.3249,1.4982) circle[radius=0.05];
    \fill (4.3249,-0.23386) circle[radius=0.05];
    \draw (4.3249,1.4982) -- (3.3249,1.4982) -- (2.6624,0.7490);
    \draw (4.3249,1.4982) -- (4.3249,-0.23386) -- (4.6424,-1.18213);
  \end{scope}
  \begin{scope}[red, thick]
    \fill (3.4599,-2.9964) circle[radius=0.05];
    \fill (4.9599,-2.1304) circle[radius=0.05];
    \fill (6.9199,0) circle[radius=0.05];
    \fill (6.9199,-1.7320) circle[radius=0.05];
    \draw (3.4599,-2.9964) -- (4.9599,-2.1304) -- (4.6424,-1.18213);
    \draw (6.9199,0) -- (6.9199,-1.7320) -- (5.9399,-1.9312) -- (4.9599,-2.1304);
  \end{scope}
\end{tikzpicture}
  \caption{Local cross-component configurations determining the lower endpoint
  of the cyclone pattern.}
  \label{fig:tri-cyclone-kissing}
\end{figure}

\end{document}